\documentstyle[12pt]{article}
\input html.sty
\begin{document}
\title{MATTERS OF GRAVITY, The newsletter of the APS Topical Group on 
Gravitation}
\begin{center}
{ \Large {\bf MATTERS OF GRAVITY}}\\
\bigskip
\hrule
\medskip
{The newsletter of the Topical Group on Gravitation of the American Physical 
Society}\\
\medskip
{\bf Number 14 \hfill Fall 1999}
\end{center}
\begin{flushleft}

\tableofcontents
\vfill
\section*{\noindent  Editor\hfill}

\medskip
Jorge Pullin\\
\smallskip
Center for Gravitational Physics and Geometry\\
The Pennsylvania State University\\
University Park, PA 16802-6300\\
Fax: (814)863-9608\\
Phone (814)863-9597\\
Internet: 
\htmladdnormallink{\protect {\tt pullin@phys.psu.edu}}
{mailto:pullin@phys.psu.edu}\\
WWW: \htmladdnormallink{\protect {\tt http://www.phys.psu.edu/\~{}pullin}}
{http://www.phys.psu.edu/\~{}pullin}
\begin{rawhtml}
<P>
<BR><HR><P>
\end{rawhtml}
\end{flushleft}
\pagebreak
\section*{Editorial}

I wanted to remind people to sign up and keep current their membership in
the Topical Group in Gravitation. Our membership was oscillating around 600,
and we need 1200 to become a division. Please talk to your friends and 
colleagues in the area about joining. 

The next newsletter is due February 1st.  If everything goes well this
newsletter should be available in the gr-qc Los Alamos archives under
number gr-qc/9909022. To retrieve it send email to 
\htmladdnormallink{gr-qc@xxx.lanl.gov}{mailto:gr-qc@xxx.lanl.gov}
(or 
\htmladdnormallink{gr-qc@babbage.sissa.it}{mailto:gr-qc@babbage.sissa.it} 
in Europe) with Subject: get 9909022
(numbers 2-8 are also available in gr-qc). All issues are available in the
WWW:\\\htmladdnormallink{\protect {\tt
http://vishnu.nirvana.phys.psu.edu/mog.html}}
{http://vishnu.nirvana.phys.psu.edu/mog.html}\\ 
A hardcopy of the newsletter is
distributed free of charge to the members of the APS
Topical Group on Gravitation. It is considered a lack of etiquette to
ask me to mail you hard copies of the newsletter unless you have
exhausted all your resources to get your copy otherwise.

If you have comments/questions/complaints about the newsletter email
me. Have fun.
\bigbreak

\hfill Jorge Pullin\vspace{-0.8cm}
\section*{Correspondents}
\begin{itemize}
\item John Friedman and Kip Thorne: Relativistic Astrophysics,
\item Raymond Laflamme: Quantum Cosmology and Related Topics
\item Gary Horowitz: Interface with Mathematical High Energy Physics and
String Theory
\item Richard Isaacson: News from NSF
\item Richard Matzner: Numerical Relativity
\item Abhay Ashtekar and Ted Newman: Mathematical Relativity
\item Bernie Schutz: News From Europe
\item Lee Smolin: Quantum Gravity
\item Cliff Will: Confrontation of Theory with Experiment
\item Peter Bender: Space Experiments
\item Riley Newman: Laboratory Experiments
\item Warren Johnson: Resonant Mass Gravitational Wave Detectors
\item Stan Whitcomb: LIGO Project
\end{itemize}
\vfill
\pagebreak

\section*{\centerline {Does the GSL imply an entropy bound?}}
\addtocontents{toc}{\protect\medskip}
\addtocontents{toc}{\bf Research briefs:}
\addtocontents{toc}{\protect\medskip}
\addcontentsline{toc}{subsubsection}{\it  
Does the GSL imply an entropy bound?, by Warren G. Anderson}
\begin{center}
    Warren G. Anderson, University of Wisconsin-Milwaukee\\
\htmladdnormallink{warren@ricci.phys.uwm.edu}
{mailto:warren@ricci.phys.uwm.edu}
\end{center}
\parindent=0pt
\parskip=5pt

Recently, a number of papers have appeared on the gr-qc preprint archive
concerning entropy bounds imposed on matter by the generalized second law of
thermodynamics (GSL) for spinning and charged black holes[1]. However, the
question of whether the GSL implies the existence of such bounds, even for the
simple case of a Schwarzschild black hole, is still being debated in the
literature. This question, first raised by Bekenstein[2] almost 30 years ago,
arises from a {\it gedankenexperiment}. The goal of this
article is to review this experiment and the key results which comprise our
current understanding of whether or not the GSL implies an entropy bound for
matter.

Let us begin by recalling the GSL: any process involving matter
with entropy $S_{matter}$ exterior to a black hole of area ${\cal A}$ should
satisfy $\delta S_{matter} + \delta S_{BH} \ge 0$, where $S_{BH}={\cal A}/4$ is
the Bekenstein-Hawking black hole entropy. The central premise of all the
papers reviewed here is that this law should be valid.

Now, consider a {\it gedankenexperiment} involving a black hole (area ${\cal
A}$) and a box of proper height $b$ and cross-sectional area $A$. Far from the
black hole, the box is filled with matter, so that the total energy of the box
and contents is $E$ and its total entropy $S$. The box is then lowered
adiabatically toward the black hole, so that its entropy $S$ remains constant.
Its energy as measured by a distant static observer, $E_\infty$, however, does
not; the box is doing work on the agent that is controlling the lowering
process. This work is subtracted from the box in the form of a redshift. When
the box is a proper height $\ell$ above the horizon, the energy in the box as
measured at infinity is therefore $E_\infty(\ell)=\chi(\ell) E$, where
$\chi(\ell)$ is the redshift factor.

Suppose that the box is lowered until it nearly touches the black hole. This
is, of course, physically impossible, since the box and/or rope will fail
mechanically before this can happen. However, for simplicity we consider this
limiting case. In this limit the box is an average proper distance $\sim b$
from the horizon, and the energy of the box is $E_\infty(b) \approx
E~b~\sqrt{\pi/{\cal A}}$.  If the box is then allowed to fall freely into the
black hole, the entropy of exterior matter will be reduced by $\delta
S_{matter} = - S$, but the entropy of the black hole will increase by $\delta
S_{BH} = 2~E_\infty(b)~\sqrt{\pi {\cal A}}$.  Putting these entropy changes
into the GSL, one gets 
$$
   S/E \le 2 \pi b. \eqno(1)
$$
This inequality, originally derived by Bekenstein[2], seems to imply a new law
of nature: that any matter of energy $E$ confined to a volume whose smallest
dimension is $b$ has a fundamental upper bound on its entropy.

The derivation of this bound, however, depends on the box continuing to do
work on the lowering agent all the way to the horizon. If this is not the
case, the bound might be modified, or even removed. About a year after
Bekenstein proposed bound (1), Unruh and Wald[3] pointed out that quantum
effects could indeed alter the work done by the box, and hence the entropy
bound.

The Unruh-Wald argument goes as follows: an adiabatic lowering process is
quasi-static, and can therefore be treated as a sequence of static (i.e., 
accelerating) boxes. Accelerating observers see the quantum vacuum as a bath
of thermal radiation, whose temperature is proportional to the
acceleration. Since the bottom of the box is closer to the black hole than the
top, it must have a greater acceleration to remain static. Thus, the bottom of
the box is exposed to hotter radiation than the top, creating a net upward
pressure. This pressure gradient buoys the box against gravity, reducing the
net work done by the box on the lowering agent. In fact, since the
acceleration (and hence temperature) diverges at the horizon, near the horizon
the lowering agent would have to do work on the box (push the box) in order to
lower it further. Thus, at some distance above the horizon, where the box
stops doing work and the lowering agent starts doing work, the box must float
freely above the black hole. Just as for Archimedes' buoyancy, the box floats
when its energy is equal to the energy of the displaced fluid (i.e., 
acceleration radiation).

For a buoyant box, Unruh and Wald[3] showed that the box contributes
the minimum entropy to the black hole when dropped from the floating point.
For macroscopic boxes, this point is very close to the black hole[4], so we do
not avoid the issue of mechanical instability, but again let us consider the
limiting case (floating box) since this will minimize the entropy gain for the
system. In this limit, the minimum change in black hole entropy is[3]
$$
   \delta S_{BH} = \frac{A}{T_{bh}}\int_0^b [\rho(l_0,y)-\rho_{ar}(l_0,y)]
      \chi(l_0+y) dy + S_{ar}, \eqno(2)
$$
where $T_{bh}$ is the Hawking temperature of the black hole,  $\rho$ and
$\rho_{ar}$ are the energy densities of the box and the acceleration radiation
respectively, $l_0$ is the proper height of the bottom of the box above the
black hole at the floating point, and $S_{ar}$ is the entropy of the 
acceleration radiation displace by the box. 

Interpretation of Eq.~(2) is not difficult. It simply states that the minimum
change in black hole energy ($S_{BH}~T_{bh}$) due to an adiabatically lowered
box is the energy of the box at the floating point ($\int \rho dV$) minus the
work done by the box against the buoyancy force of the acceleration radiation
during lowering ($T_{bh} S_{ar} - E_{ar}$). This interpretation seems
reasonable, and until recently Eq.~(2) has been accepted as correct in the
literature (I will deal with a recent exception a bit later in this article).

However, the conclusions that can be drawn from Eq.~(2) regarding entropy
bounds seem to depend heavily on the nature of the acceleration radiation
(i.e., on $\rho_{ar}$ and $S_{ar}$). In particular, the crucial assumption
seems to be the following: since acceleration radiation is thermal, it should
have the maximum entropy density at any given energy density. If this is the
case, it has been shown that Eq.~(2) implies the GSL without assuming a bound
such as Eq.~(1)[3,5,6] (although Eq.~(1) itself may follow from this
assumption together with some other plausible assumptions about the
acceleration radiation[6]). Conversely, in those papers[2,4,7] where this
maximal entropy density assumption is not made, the GSL is shown to be
violated unless an entropy bound exists.

That the resolution of the entropy bound question rests on properties of
acceleration radiation is slightly troubling, because in some senses this
radiation is not physical. For instance, even though the vacuum expectation
value of the stress-energy tensor for an electromagnetic field in Minkowski
space vanishes, accelerating observers see that vacuum as a thermal bath of
photons. However, these photons clearly carry no energy or momentum on
average. One must therefore be careful as to the properties one requires of
acceleration radiation.

Recently, Bekenstein[8] has placed even more emphasis on the nature of
acceleration radiation.  He has pointed out that since the temperature far
from the black hole is low, the average wavelength of the radiation can be
longer than the height of the box. Therefore, he argues, the acceleration
radiation will not behave like a fluid there; rather, one should treat the
interaction of the radiation with the box as a scattering process. This can
significantly reduce the buoyancy, causing Eq.~(2) to be modified. He then
finds that an entropy bound of the form (1) is necessary to preserve the GSL,
even if the acceleration radiation is assumed to be maximally entropic.

If it is troubling to invoke the properties of acceleration radiation in
resolving questions about this {\it gedankenexperiment}, one might ask how
such questions can be resolved in an observer independent way. Such a
resolution was originally provided by Unruh and Wald[3], and somewhat
elucidated later[9], although it has not played a part in the recent
literature. The resolution lies in the fact that accelerating surfaces emit
quantum fluxes. These fluxes can have negative energy, and it is just such a
flux that is created inside the box due to its accelerating walls. As
the box is lowered the energy in the box decreases, since the {\it negative}
energy deposited in the box by the quantum flux is added to the energy density
of the box itself. This negative energy is just what one would have to add to
the acceleration radiation in order that an initially empty box, lowered
adiabatically toward the black hole, should continue to look empty to an
observer accelerating with the box. In other words, it preserves the adiabatic
vacuum (Boulware) state inside the box. In this picture, the box floats because
the negative energy of the accelerating vacuum inside the box cancels the 
positive energy of the box itself. 

Most interestingly, Eq.~(2) can be derived by the quantum flux analysis as
well. One might suspect, therefore, that this analysis needs to be modified so
as not to be at odds with Bekenstein's calculation of the the scattering of
acceleration radiation at low temperatures[8]. A natural modification which
might be analogous to the scattering process would be the exclusion of long
wavelength components of the flux due to the scale set by the size of the box.
However, at least in two-dimensional examples, it can been shown that Eq.~(2)
follows from an exact analysis, regardless of any exclusion of long
wavelength components[3]. The question of understanding Bekenstein's
modification to (2) without invoking acceleration radiation therefore seems to
be open. It is open questions such as this that will need to be resolved
before we can truly understand whether or not the generalized second law of
thermodynamics implies an entropy bound.

{\bf References}
\parskip=1pt

{\rm [1]} T. Shimomura and S. Mukohyama,
\htmladdnormallink{gr-qc/9906047}{http://xxx.lanl.gov/abs/gr-qc/9906047}; 
A. E. Mayo, 
\htmladdnormallink{gr-qc/9905007}{http://xxx.lanl.gov/abs/gr-qc/9905007}; 
B. Linet, 
\htmladdnormallink{gr-qc/9905007}{http://xxx.lanl.gov/abs/gr-qc/9905007}; 
S. Hod, 
\htmladdnormallink{gr-qc/9903010}{http://xxx.lanl.gov/abs/gr-qc/9903010}, 
\htmladdnormallink{gr-qc/9903011}{http://xxx.lanl.gov/abs/gr-qc/9903011}; 
J. D. Bekenstein and A. E. Mayo, 
\htmladdnormallink{gr-qc/9903002}{http://xxx.lanl.gov/abs/gr-qc/9903002}; 
S. Hod, 
\htmladdnormallink{gr-qc/9901035}{http://xxx.lanl.gov/abs/gr-qc/9901035}. 

{\rm [2]} J. D. Bekenstein, Phys. Rev. {\bf D23}, 287-297, (1981).

{\rm [3]} W. G. Unruh and R. M. Wald, Phys. Rev. {\bf D25}, 942-958, 
(1982).

{\rm [4]} J. D. Bekenstein, Phys. Rev. {\bf D49}, 1912-1921, (1994).

{\rm [5]} W. G. Unruh and R. M Wald, Phys. Rev. {\bf D27}, 2271-2276, (1983).

{\rm [6]} M. A. Pelath and R. M. Wald, 
\htmladdnormallink{gr-qc/9901032}{http://xxx.lanl.gov/abs/gr-qc/9901032}.

{\rm [7]} J. D. Bekenstein, Phys. Rev. {\bf D27}, 2262-2270, (1983).

{\rm [8]} J. D. Bekenstein, 
\htmladdnormallink{gr-qc/9906058}{http://xxx.lanl.gov/abs/gr-qc/9906058}. 

{\rm [9]} W. G. Anderson, Phys. Rev. {\bf D50}, 4786-4790, (1994).

\parskip=5pt
\newpage

\section*{\centerline {
A lightweight review of middleweight black holes}}
\addtocontents{toc}{\protect\smallskip}
\addcontentsline{toc}{subsubsection}{\it  
A lightweight review of middleweight black holes, by Ben Bromley}
\begin{center}
Ben Bromley, University of Utah\\
\htmladdnormallink{bromley@physics.utah.edu}
{mailto:bromley@physics.utah.edu}
\end{center}
\parindent=0pt
\parskip=5pt

\def\msol{M$_\odot$}

The observed distribution of black hole masses has a well-known
gap. On the one hand we have ``black hole candidates'' in our own
Galaxy with inferred masses between 3 and 10\,\msol, and on the other,
there are supermassive black holes which lie in the centers of many,
if not most, galaxies with mass in the range of $10^6$ to
$10^9$\,\msol. The mass gap would be filled by ``middleweights'' of a
thousand \msol, give or take a couple of orders of magnitude.

There are several reasonable ways to form middleweight black holes,
including merging of solar-mass black holes in star clusters (Lee
1995), or even more directly from gravitational collapse of density
fluctuations in the primordial ooze (Haiman, Thoul \& Loeb 1996),
although these mechanisms can be suppressed to some degree by
supernova-driven ejection of matter. Yet until recently, there was
virtually no evidence for the existence of middleweights.  This might
just be a selection effect: We find black hole candidates in X-ray
binary systems because they are local and have a steady source of
luminous matter flowing from a companion star. Supermassive
holes, the central engines in active galactic nuclei, are observed
because they are very bright, feeding off of the gas-rich environments
in the centers of galaxies. In contrast, it is unclear what, if any,
reservoirs of luminous matter might surround middleweights.

Within the past year, the observational situation changed
dramatically. Two groups, Ptak \& Griffiths (1999) and Colbert \&
Mushotzky (1999), reported data which suggest the presence of
middleweight black holes in nearby galaxies. Ptak \& Griffith used the
ASCA satellite to measure the hard X-ray spectrum of the starburst
galaxy M82. A compact X-ray source detected in the central region of
the galaxy has a luminosity which is two orders of magnitude larger
than the brightest black hole candidate, and two orders of magnitude
dimmer than a typical active galactic nucleus. Assuming that the
accretion flow onto the source is at the Eddington limit, the observed
luminosity corresponds to a mass of 460\,\msol.

The Eddington limit does rule out solar-mass black holes, however, it
is plausible that the X-ray source is just an unusually dim active
galactic nucleus harboring a supermassive hole.  Future high angular
resolution observations of soft X-rays from an accretion disk around
this source (by the Chandra X-ray Observatory, for example) might help
distinguish a middleweight from an ``ordinary'' supermassive hole.

Colbert \& Mushotzky used a combination of ROSAT X-ray imaging data
and ASCA spectroscopy to examine X-ray sources in 21 nearby
galaxies. They determined that a half dozen or so objects in the
sample show X-ray emission from compact sources well away from the
photometric center of their host galaxies. This adds some spice into
the mix since some of these sources also have high luminosities and
hence high Eddington masses. A suggestion that these sources are just
unusually quiescent active galactic nuclei is significant, since it
would be interesting to observe nuclear activity well outside the
galactic nucleus.

Of the 21 objects in the sample, ASCA yielded hard X-ray spectra from
three spiral galaxies and three ellipticals. The X-ray sources in the
ellipticals are all coincident with the photometric centers (within
positional errors). While their spectra show features similar to
Galactic X-ray binaries, the luminosities are more than an order of
magnitude higher, suggesting at least middleweight black
holes. However, low-luminosity active galactic nuclei should not be
ruled out. In two of the three ellipticals, Hubble Space Telescope
observations of circumnuclear disks indicate the presence of
supermassive black holes of more than $10^8$\,\msol. We may not be
able to say much, it seems, from the X-ray flux alone.

The spirals discussed by Colbert \& Mushotzky perhaps make for better
candidates as hosts of middleweight black holes. Two of them, NCG~1313
and NGC~5408, have an extranuclear X-ray source that is roughly a
kiloparsec away from the photometric center of the galaxy.  These
objects have X-ray fluxes which are also about an order of magnitude
greater than expected from known X-ray binaries containing a black
hole candidate. The lack of any other evidence for behavior typical of
active galactic nuclei, plus the dynamical awkwardness of placing a
supermassive black hole in the suburbs of otherwise normal galaxies
lends some credence to the idea that the sources are middleweights.

What could be wrong with the middleweight interpretation? The
Eddington luminosities suggest mass limits of roughly 10\,\msol, hence
the sources might only be pushing the boundary of what we now consider
normal black hole candidates of stellar origin (e.g., Fryer 1999). Other
possibilities include superluminous sources from which Eddington
limits are not useful, and X-ray supernovae which can remain at
constant brightness for periods of years. However, sub-Eddington
accretion onto a middleweight is reasonable, and with luminosities at
a percent or even a tenth of the Eddington limit, the mass of the
black holes in NCG~1313 and NGC~5408 becomes interesting.

A middleweight black hole sounds at first like a gift from the heavens
for the earth based gravity wave detectors.  The frequency of peak
sensitivity for these detectors corresponds to strong field processes
for holes of a hundred or so \msol~(Flanagan \& Hughes 1998). But
middleweights need more than mere existence to be the most fascinating
objects in the gravitational night sky. The middleweights must also
take part in some process that generates strong waves. The inspiral
and plunge of a stellar mass compact object (hole or neutron star)
would produce waves of the right frequency, but relatively low
amplitude. What would be ideal would be the merger of two
middleweights, perhaps as a step toward the formation of a
supermassive hole. Whether this is a plausible astrophysical scenario,
or a gift that the heavens won't deliver, depends on what sort of
processes produce the middleweights, and what astrophysical
neighborhoods they inhabit.  These questions will be studied in the
coming year, but near term answers are likely to be very speculative.

\noindent
{\bf References}

\noindent
Colbert, E. J. M., \& Mushotzky, R. F. 1999, 519, 89 \\
Flanagan, \'E. \'E,  \& Hughes, S. 1998, Phys. Rev. D, {57}, 
4535 (also \htmladdnormallink{gr-qc/9701039}
{http://xxx.lanl.gov/abs/gr-qc/9701039}) \\
Fryer, C. 1999, preprint (
\htmladdnormallink{astro-ph/9902315}
{http://xxx.lanl.gov/abs/astro-ph/9902315}
) \\
Haiman, Z., Thoul, A. A., \& Loeb, A. 1996, ApJ 464, 523 \\
Lee, H. M. 1995, MNRAS, 272, 605 \\
Ptak, A. \& Griffiths, R. 1999, ApJ, 517, L85 \\

\newpage

\section*{\centerline {
The physics of isolated horizons}}
\addtocontents{toc}{\protect\smallskip}
\addcontentsline{toc}{subsubsection}{\it  
The physics of isolated horizons, by Daniel Sudarsky}
\begin{center}
Daniel Sudarsky, ICN-UNAM, Mexico\\
\htmladdnormallink{sudarsky@nuclecu.unam.mx}
{mailto:sudarsky@nuclecu.unam.mx}
\end{center}
\parindent=0pt
\parskip=5pt

  In the last couple of years there has been substantial progress in
 various paths toward the elucidation of the deep relationship that
 emerges in the study of the classical dynamics of black holes, the
 behavior of quantum fields in background black hole spacetimes, and
 the ordinary laws of thermodynamics.  One of the most successful is
 the String Theory program which has produced a detailed evaluation of
 the statistical mechanical entropy of some extremal and near extremal
 stationary black holes, through the relation of these objects with a
 certain class of states in the weak gravity sector of the
 theory. This last feature is what makes the approach somehow
 unsatisfying to some researchers approaching the question from the
 gravity point of view, who would like to know, for example, where do
 the degrees of freedom that account for the black hole entropy
 reside?  i.e., the horizon's surface? its exterior? its interior?.

 The second
 program that has met recently with substantial success is the
nonperturbative quantum geometry program also known as
"loop quantum gravity" [1].The first step has been
to both, generalize and
 properly define the sector of the theory that is going
 to be treated. In doing so, Ashtekar and his colleagues [3][4], were
guided
 by the need
to start with a well defined action that would be differentiable
 in the sector under consideration. This leads to the specialization
 of the notion of Trapping Horizons [2] of Hayward's to that of
Isolated Horizons  [3] . Physically the idea is to represent
"horizons in internal equilibrium and decoupled from what is outside".
 Essentially, an isolated horizon is defined to be a null 3 surface
$\Delta$, with topology $S^2 \times R $, with nonexpanding null
tangent vector
 field $l^a$ which is also a Killing field for the induced
metric on  $\Delta$, foliated by a preferred set of 2-spheres
$\lbrace S \rbrace $ that are marginally trapped, and such that the
induced metric
 on each $S$ is spherically symmetric. The definition includes also the
 requirement that the horizon be nonrotating which demands among
other things that the second null vector field that is normal to
 the 2-surfaces $S$ of the isolated horizon $n^a$, be shear free and
have a spherically symmetric expansion. Moreover, in defining the class of
 spacetimes containing isolated horizons one requires that the field
equations be satisfied on the horizon, (not so elsewhere in the
spacetime).

 Here, we must point out that the
 spacetimes themselves are not assumed to have any Killing field, not
even in a neighborhood of the horizon, and as such the class is
extremely large, in contrast, say, with the 3 parameter class of
stationary black hole solutions of Einstein Maxwell theory, as
they include for example black hole spacetimes with nonstationary
 matter or gravitational waves in the exterior (in these cases the
 horizon will be isolated for as long as those nonstationary components
have not crossed the horizon). Moreover, given that the definition of
isolated horizon is semilocal and does not rely, for example,
in the existence of asymptotic null infinity, the class of
spacetimes containing isolated horizons is not limited to the
 asymptotically flat case and includes for example cosmological
 examples such as de Sitter spacetime. The crucial point of the
definition is that it is possible to add a surface term to the
usual bulk action of general relativity, possibly coupled with
 suitable matter fields like Maxwell and scalar fields such that
the action is differentiable within the corresponding class of
 spacetimes. This is analogous to the usual addition of a surface
 term to the bulk action of general relativity on manifolds with
boundary formulated in terms of the spacetime metric in such a way
 that it becomes differentiable on the class of  spacetimes with a
 fixed value of the metric on  the manifold boundary. This is a very
 important, and nontrivial point for the quantization program,
because one needs to start with a classical action and configuration
 space from which one can extract not only the equations of motion
but also the symplectic form. This can be achieved through a well
defined procedure, once such a differentiable action
is provided [5]. The surface term that must be added when the spacetimes
 under consideration are taken to have an isolated horizon as one of
 its boundaries and when  the gravitational variables are taken to be
  the soldering form and the  spin connection, turns out to be the
Chern Simons action for a $U(1)$ connection on $\Delta$. The term
 itself depends on the value of the horizon area, thus the action
 is appropriate for a class of spacetimes with fixed horizon area $A$.

 Obviously, if the class of spacetimes under consideration allows other
 boundaries, one must add the corresponding surface terms to ensure
 differentiability of the action within the class. The fact that
such a differentiable action exists seems to depend completely on
 the choice of the physical boundary conditions and not so much on
 the choice of variables, as long as the action is a first order
action as in the Palatini formulation.
\medskip

 Equipped with the above structure the procedure to get
 a Hamiltonian Formulation is straight forward, i.e.,
consider a foliation of spacetime (which is assumed to
have the topology $\Sigma \times R$), introduce lapse and shift
 and identify the canonical variables, which in this case
include, in addition to the ordinary Hamiltonian variables
 of bulk gravitational sector and possibly the matter fields,
the projection of the $U(1)$ connection on the intersection on
 the isolated horizon with the hypersurface $\Sigma$. Thus one has
 an adequate formulation corresponding to a phase space $\Gamma_A$
associated with configurations with a fixed value $A$ for the area
 of the isolated horizon.

 The stage is set to look at the  "thermodynamics"
 of these isolated horizons. The first step is to
 define a notion of surface gravity which at first sight seems
to be straight forward given the existence of a null Killing field
for the metric of the isolated horizon, however one immediately
 faces the problem of choice of normalization for this vector field.
In the case of stationary black holes, this task is accomplished
 through the normalization of the Killing fields at asymptotic
infinity, and thus, the same strategy is unavailable in the
isolated horizon case, because no such Killing field is in
 general available for the whole spacetime (in general, there
 is not even an asymptotically flat boundary). The problem is
 solved by fixing the expansion $n^a$ to coincide with the value it
 takes in Reissner Nordstrom, and then fixing the normalization of
$l^a$ by $l^a n_a =-1$. It is noteworthy fact that such a simple recipe
 exists which results in the correct value for the surface gravity of
 the static black holes in Einstein Maxwell Dilaton Theory.

  Not so surprisingly, given the high degree of symmetry of the
horizon itself, the surface gravity thus defined turns out to be
constant on the isolated horizon. Therefore, the zeroth law of
 "thermodynamics" of isolated horizons holds. Nevertheless, we
 must note that the identification of the surface gravity of
general isolated horizons with a physical temperature is not clear
since there is so far no analog to the analysis that established
 the phenomena of Hawking radiation.

  The next step is to define a notion of mass associated with
the isolated horizon, which given the general absence of an
asymptotically flat boundary can not be taken to be the ADM mass,
 and, moreover, even when there is such a boundary the fact that
there is  general matter and gravitational waves in the exterior
 spacetime the ADM mass would depend on those fields and not only
on the isolated horizon itself. Notably the answer lies in the
construction of an appropriate hamiltonian, i.e. one that
 would give the correct equations of motion upon considerations
of arbitrary variations within the phase space (i.e. variations
that not necessarily vanished at the boundaries and, in particular,
 at the isolated horizon). Moreover, it will be necessary to consider
a new phase space $\Gamma'$ constructed by taking the union of the
 phase spaces of isolated horizons for all possible values of the
 horizon area. This requires the addition of a surface term associated
 with the isolated horizon boundary, and a choice of lapse and shift
corresponding to l/a at the isolated horizon boundary, that
 is completely analogous to the usual addition of the ADM
mass term in association with asymptotically flat boundaries
 and  evolution with a choice of lapse and shift that correspond
to a time translation at infinity. The fact that such a boundary
 term making the Hamiltonian differentiable exists is highly nontrivial,
 in particular, no such term is known to exist in the case of general
internal boundaries with general choices of lapse and shift.
This boundary term in the Hamiltonian is naturally identified
with the mass of the isolated horizon and coincides with the ADM mass
in the case of static black holes in Einstein Maxwel Dilaton theory.

 One is then in the position of considering the first
law of thermodynamics of Isolated Horizons. A straight
forward calculation yields
 $$
  \delta M_\Delta = {\kappa \over {8 \pi}} \delta A + work terms
 $$
 Thus establishing the validity of the first law.  We note that the
physical process version of the first law is not fully satisfactory
because, strictly
speaking, the intermediate stages of the process need not contain isolated
 horizons, and, therefore, do not correspond to points in $\Gamma'$
Related concerns can be raised about the usual analysis,
as well. We note, for example, that
through a physical process the ADM mass can not change, so,
 in this last respect, the
Isolated Horizon approach seems more satisfactory.

 There is, at present, no analog to the second law for isolated horizons.
 This is due in part to a problem similar to that mentioned above, namely
 the fact that in the definition of the notion of isolated horizons one
leaves no room for a situation in which the area of the horizon changes.
 Furthermore, in this case the problem can not be sidestepped even through
an approximation because the second law is supposed to have a validity
that goes quite beyond the quasistationary regime.

 Finally, the program turns out to be very
 successful in evaluating the statistical mechanical
 value of the entropy of an isolated horizon. This part
starts with the quantization of the appropriate sector,
namely the theory associated with the phase space with the
 standard Ashtekar bulk variables on an hypersurface $\Sigma$
 with boundary S corresponding to the isolated horizon,
together with the $U(1)$ connection of the
Chern Simons theory on the boundary. The theory is
 as usual, subject to the constraints which in  this
case involve the not only the Hamiltonian, diffeomorphism and
Gauss constraints, but an additional one, inherited from
the conditions defining  the isolated horizon, that links
 the behavior on the $U(1)$ connection on S with that of
the soldering form in the bulk, evaluated at S. Having
 constructed the quantum version of the sector of isolated
 horizons one looks at the states corresponding to eigenvalues
 of the area operator for the boundary $ S $
(the area of the isolated horizon) lying within the range $A-l_P^2$ and
$A + l_P^2$. Then, one takes the maximal entropy density matrix
 made out of these states, namely the equally weighed
totally uncorrelated density matrix constructed from these
 states, and construct the density matrix describing the
 horizon degrees of freedom by tracing over the bulk degrees
 of freedom. Then, the evaluation of the entropy through the
standard formula $-tr \rho \ln \rho$ yields, in the case of a large
 enough horizon area, the result to lowest order is
 $$
  S = { \ln 2 \over  {4 \pi \sqrt 3 l_P^2}}  \gamma A
 $$
 Thus, by selecting the value $ \gamma = \ln 2 / (\pi \sqrt 3 )$
for the Immirzi parameter $\gamma$,
(which amounts to selecting one  among a
continuous choice of unitarily unequivalent
quantum theories corresponding to the same classical theory),
the standard result $ S = A/4l_P^2 $ is obtained. It is worth to
point out that this choice can be made only once, and that the
 number of different situations that must be accounted by such
 a choice is infinite so it is a highly nontrivial fact that
 such a choice exits. For example, it is conceivable that say
 the choice needed in the case of Reissner Nordstrom black holes
would have been different than the choice needed for
Schwarzschild black holes.

 The program has therefore met with tremendous success so
 far, and the task of generalizing the setting to account
 also for rotating black holes is currently under way. More
into the future, there is the hope of eventually treating
fully dynamical horizons,  and of replacing the effective analysis
 described above with a more fundamental one, in which, starting
 with the full quantum theory one could single out the appropriate
 sector of states and carry out the appropriate counting for
the evaluation from first principles of the horizon entropy.
  We look forward to these and other developments resulting from
 this exciting program, as well as for new insights from other
 sources, in the hope that eventually we would be in a position
 to treat even more vexing questions such as those related to
the ultimate fate of an evaporating black hole and the issue
of information loss.

{\bf References:}

 [1] A. Ashtekar, Quantum Mechanics of Geometry,
The Narlikar Festschrift, ed. N. Dadhich and A. Kemhavi 1999,
\htmladdnormallink{gr-qc/990123}{http://xxx.lanl.gov/abs/gr-qc/990123}.
C. Rovelli, "Loop Quantum Gravity", Living reviews in Relativity, No.
1998-1. \htmladdnormallink{gr-qc/9710008}
{http://xxx.lanl.gov/abs/gr-qc/9710008}.
Ashtekar, Rovelli, Smolin, "Weaving a classical geometry with quantum
threads",
Phys.Rev.Lett.69:237-240,1992

 [2] S. Hayward, General laws of black hole dynamics,
Phys. Rev. D {\bf 49}, 6467 (1994).

 [3]A. Ashtekar, A. Corichi and K. Krasnov, Isolated Horizons:
The classical phase space, Adv. Theor. Math. Phys., in press,
\htmladdnormallink{gr-qc/9905089}{http://xxx.lanl.gov/abs/gr-qc/9905089}.
 A Ashtekar, C. Beetle and S. Fairhurst, Mechanics of
 Isolated Horizons 
\htmladdnormallink{gr-qc/9907068}{http://xxx.lanl.gov/abs/gr-qc/9907068}

 [4] A Ashtekar, J. Baez, A. Corichi and K Krasnov, Quantum Geometry and
Black
 Hole Entropy, Phys Rev. Lett {\bf 80}, 904 (1998).

 [5] See for instance  Sec. II in
 J. Lee and R. Wald, Local Symmetries and Constraints, J. Math Phys.
{\bf 31}, 725, (1990).

\newpage

\section*{\centerline {LIGO project update}}
\addtocontents{toc}{\protect\smallskip}
\addcontentsline{toc}{subsubsection}{\it  
LIGO project update, by Stan Whitcomb}
\begin{center}
Stan Whitcomb, Caltech\\
\htmladdnormallink{stan@ligo.caltech.edu}
{mailto:stan@ligo.caltech.edu}
\end{center}
\parindent=0pt
\parskip=5pt

Work on the Laser Interferometer Gravitational-wave Observatory
(LIGO) continues to progress smoothly. The bake out of the beamtubes at
the Hanford site was concluded in May, completing the facilities and
infrastructure work there.  The bakeout equipment have now been shipped to
the Livingston site, and the bakeout of the first beamtube module should
start within weeks.

The main activities have now shifted to detector installation at both
sites.  At Hanford, the seismic isolation installation for the first
interferometer is nearly complete, and over half of the optics have been
installed in the vacuum chambers.  The prestabilized laser has been
installed and under test for several months.  A major current focus is
locking the laser to the mode cleaner, the first step in  integrating and
commisioning the first detector.  At Livingston, the first installations
are the seismic isolation system and the prestabilized laser; both are
well underway.

At the most recent meeting of the LIGO Scientific Collaoration, held at
Stanford in July, the main focus was initial discussion and planning for
a improved detectors to replace the current ones starting in 2005.  Although
there is significant R\&D to completely define these detectors, it is clear
that significant sensitivity improvements are possible.

Progress on LIGO (including pictures of the installation activities)
can be followed in our (nearly) monthly newsletters accessible through
our website
(\htmladdnormallink{http//:www.ligo.caltech.edu}{http//:www.ligo.caltech.edu}).

\newpage

\section*{\centerline {Worskhop on initial value for binary black holes}}
\addtocontents{toc}{\protect\medskip}
\addtocontents{toc}{\bf Meeting reports:}
\addtocontents{toc}{\protect\medskip}
\addcontentsline{toc}{subsubsection}{\it  
Worskhop on initial value for binary black holes, by Carlos Lousto}
\begin{center}
Carlos Lousto, Albert Einstein Institute, Golm, Germany\\
\htmladdnormallink{lousto@aei-potsdam.mpg.de}
{mailto:lousto@aei-potsdam.mpg.de}
\end{center}
\parindent=0pt
\parskip=5pt

The first (initial) workshop on initial data for binary black holes
took place at the new location of the AEI: Golm, 
June 7-9 1999. It was extremely successful regarding the very high level
of attendance and participation. Copies of the related papers (when available)
prior to the meeting have been distributed as ``precedings'', and 
a certain flexibility in the schedule provided a fertile atmosphere for
numerous questions and discussion.

The first day emphasized studies within the conformally flat
ansatz for the 3-geometry. G. Cook presented a comprehensive review of
the classical results and introduced new developments like
the ``thin sandwich approach'' recently proposed by York[1].
He noted how this approach turns out to naturally include the approximations
used by Mathews and Wilson and the ``convective'' one used by Lousto
and Price in the particle limit.  Cook also emphasized the need to move
beyond the conformally flat ansatz and Bowen-York solutions of the momentum
constraints in the quest to find more astrophysically realistic initial data.
He also expressed interest in using the ``lambda'' systems described by
Brodbeck, Frittelli, H\"ubner, and Reula for enforcing the constraints 
during numerical evolutions.

B. Br\"ugmann presented the black hole punctures approach to initial data
based on Ref.\ [2]. Here the wormhole topology of black hole data is
compactified to $R^3$, which leads to an existence and uniqueness proof
and facilitates numerical studies.

C. Lousto described work with R. Price to test the Bowen-York initial
data in the black hole plus a particle system, which can be treated
perturbatively. The conclusion is that the ``longitudinal'' ansatz for
the extrinsic curvature seems not to be a good representation of an
astrophysically realistic scenario [3]. It was noted the interest on
extending these studies from headon to generic orbits.

R. Beig described a formulation of the momentum constraints where
quantities such as ADM linear and angular momentum are encoded
by specific source terms in these equations concentrated at the punctures
arising from conformally compactifying spatial infinity. He showed a
way of writing down, in the conformally flat case, solutions for these
inhomogeneous equations which are of the "Bowen-York"- type. One can then
find the general solution, since the methods of Ref.\ [4] enable one to write
down the general "unpunctured" TT-tensor. 

The second day of the workshop started with an excellent review of the
close limit approach to the final merger stage of two black holes by
J. Pullin. He also discussed the last results on perturbations of black
holes with angular momentum [5].

E. Seidel gave a comprehensive overview of the projects
carried out in the numerical group at AEI. He described the results
obtained by evolving (with the CACTUS code) black hole
plus brill wave and pure brill wave data. For this latter data it is
possible to follow either the scattering of waves leaving 
back flat spacetime or their collapse to form a black hole and then to
obtain the corresponding quasinormal ringing.

M. Campanelli introduced the ``Lazarus project'' that proposes to
marry full numerical techniques to describe well detached black holes with
perturbative techniques to continue the evolution once a common horizon
encompasses the binary system. The matching is performed by constructing
the Weyl scalars on the perturbative slice [6], and evolve them using
the Teukolsky equation.

C. Lousto then described the evolution of ``exact'' Misner data for two
initially at rest black holes via the Zerilli and the Teukolsky equations.
The results show that while the evolution with the Zerilli equation suffers
a premature break down, the evolution with the Teukolsky equation is robust
and generates results close to the linear initial data [7]. It Remains open
the question whether this behavior also holds for non headon collisions.

R. Price neatly described his solution to the constraints representing
two Kerr black holes in an axially symmetric configuration. He used the
black hole plus brill wave form of the metric and superposed exact
``Kerr'' solutions to the momentum constraint. When one imposes this
family of solutions, parametrized by the separations of the holes, to have
a close Kerr limit, a ``pin pole'' between the holes appears [8].
Whether this is a generic feature
or a consequence of the restrictive ansatz remains an open question.

W. Krivan showed the results of the evolution of the above initial data in
the close limit regime where the ``pin pole'' is safely enclosed by the
common horizon [9].

J. Baker discussed a similar approach to the problem, but he assumes the
superposition of the holes at the level of the 3-metric and then solves
for the extrinsic curvature [10]. Again, the requirement of the holes being
kerr-like both when well separated and when close together generates
unpleasant features such as discontinuities in the extrinsic curvature.

N. Bishop carefully described the Kerr-Schild ansatz to the initial
value problem [11].
This is a quite original approach and has interesting potentialities.
One needs still to identify the physical parameters and give explicit
solutions for rotating black holes. He also mentioned the related
Matzner et al work where two Kerr black holes are explicitly superimposed.

J. Thornburg [12] presented numerical initial data on an Eddington-Finkelstein
slice of the spacetime with nonvanishing K. He stressed the benefits of working
with 4th order evolution codes.

H. Shinkai exposed his work on how the post-Newtonian expansion can be
used to give initial data for a further general relativistic evolution [13].
He applied this approach to binary neutron stars and measured the precision
of several PN orders using as a criteria the violation of the Hamiltonian
constraint. For the first and second PN expansion he gets around 60\% and 45\%
errors respectively.

G. Nagy discussed about the differentiability of the Cauchy data
by studying the constraints in the case of incompressible
models of neutron stars. He proved rigorously its $C^1$ behavior at the
boundary. Extension to more realistic equations of state is being undertaken.

P. H\"ubner presented a review of the conformal approach to GR and highlighted
the stable numerical implementation of his evolution code (4th order)
due to the explicit first order symmetric hyperbolic formulation [14].
Work is in progress on the question of giving astrophysical initial data and
including matter sources.

On Thursday 10th we had two round tables on further work on initial data
and other works on subjects related to gravitational radiation.
This was planned to finish by 1 p.m. but given the interest of the
participants it was extended to the whole afternoon with discussion of
new ideas and even some computations!

Details and the full workshop program can be found in\\
\htmladdnormallink{http://www.aei-potsdam.mpg.de/\~{}lousto/WID99.html}
{http://www.aei-potsdam.mpg.de/\~{}lousto/WID99.html}.

\newpage
\parindent=0pt

{\bf References}

[1] J.W.~York,
``New data for the initial value problem of general relativity,"
\htmladdnormallink{gr-qc/9810051}{http://xxx.lanl.gov/abs/gr-qc/9810051}.

[2] S.~Brandt and B.~Br\"ugmann,
``A simple construction of initial data for multiple black holes,"
Phys. Rev. Lett. {\bf 78}, 3606 (1997)

[3] C.O.~Lousto and R.H.~Price,
``Improved initial data for black hole collisions,"
Phys. Rev. {\bf D57}, 1073 (1998)

[4] R.~Beig,"TT-tensors and conformally flat structures on 3-manifolds",
in: Mathematics of Gravitation, Part 1, Lorentzian Geometry and Einstein
Equations (P.T. Chrusciel, Ed.), Banach Center Publications {\bf 41}, 109
(1997), also: 
\htmladdnormallink{gr-qc 9606055}{http://xxx.lanl.gov/abs/gr-qc 9606055}.

[5] G.~Khanna {\it et al.},
``Inspiralling black holes: The Close limit,"
\htmladdnormallink{gr-qc/9905081}{http://xxx.lanl.gov/abs/gr-qc/9905081}.

[6] M.~Campanelli, C.O.~Lousto, J.~Baker, G.~Khanna and J.~Pullin,
``The Imposition of Cauchy data to the Teukolsky equation. 3.
The Rotating case,"
Phys. Rev. {\bf D58}, 084019 (1998)

[7] C.O.~Lousto ``Linear evolution of nonlinear initial data for binary
black holes: Zerilli vs. Teukolsky'', AEI-1999-7. 

[8] W.~Krivan and R.H.~Price,
``Initial data for superposed rotating black holes,"
Phys. Rev. {\bf D58}, 104003 (1998).

[9] W.~Krivan and R.H.~Price,
``Formation of a rotating hole from a close limit headon collision,"
Phys. Rev. Lett. {\bf 82}, 1358 (1999)

[10] J.~Baker and R.S.~Puzio,
``A New method for solving the initial value problem with application to
multiple black holes," Phys. Rev. {\bf D59}, 044030 (1999)

[11] N.T.~Bishop, R.~Isaacson, M.~Maharaj and J.~Winicour,
``Black hole data via a Kerr-Schild approach,"
Phys. Rev. {\bf D57}, 6113 (1998)

[12] J.~Thornburg,
``Initial data for dynamic black hole space-times in (3+1) numerical
relativity," Phys. Rev. {\bf D59}, 104007 (1999). 
\htmladdnormallink{gr-qc/9801087}{http://xxx.lanl.gov/abs/gr-qc/9801087}.

[13] H.~Shinkai, ``Truncated postNewtonian neutron star model," 
\htmladdnormallink{gr-qc/9807008}{http://xxx.lanl.gov/abs/gr-qc/9807008}.

[14] P.~H\"ubner, ``A Scheme to numerically evolve data for the
conformal Einstein equation," 
\htmladdnormallink{gr-qc/9903088}{http://xxx.lanl.gov/abs/gr-qc/9903088}.

\newpage

\section*{\centerline {
ITP Conference on strong gravitational fields}}
\addtocontents{toc}{\protect\smallskip}
\addcontentsline{toc}{subsubsection}{\it  
ITP Conference on strong gravitational fields, by Don Marolf}
\begin{center}
Don Marolf, Syracuse University\\
\htmladdnormallink{marolf@suhep.phy.syr.edu}
{mailto:marolf@suhep.phy.syr.edu}
\end{center}
\parindent=0pt
\parskip=5pt

The ITP Conference on Strong Gravitational Fields (Santa Barbara, CA,
June 22-26, 1999) was the culminating event of the six month
ITP program ``Classical and Quantum Physics of Strong Gravitational Fields.''
The intent of the conference was to focus on those aspects of
strong gravitational fields that had been emphasized in the six month
program, but to extend the audience far beyond the program participants.

One of the most unusual aspects of the conference was the basic
organization.  This was to be a ``Discussion Conference.'' The
schedule featured a relatively small number of invited talks, each
of which was followed by 40 minutes of discussion time.  This time
allowed time for both an in depth conversation between speaker and audience (and
often among the audience members) as well as for short presentations
by various participants on topics related to the main talk.  I think that
everyone was impressed at just how well this format seemed to work and
the extend to which the discussions clarified and brought forth important
issues from the talks.  I was especially told by the less experienced
participants that this format allowed them to see more clearly what
were the important parts of a given talk.  The conference
also featured a few (14) contributed presentations, and a panel
discussion (Bob Wald, J\"urgen Ehlers, Pablo Laguna, and Beverly Berger)
on classical relativity.  The full audio recordings
of all talks, short presentations, and ensuing discussions are available
on line at the above URL.

Like the preceding ITP program, the conference had three main foci.  One of these
was quantum gravity, especially loop quantum gravity and string theory.
Review talks on these subjects were given by Gary Horowitz, Joe Polchinski,
Rob Myers, Abhay Ashtekar, John Baez, and Carlo Rovelli.  Another
was classical physics, including Gravitational waves (John Friedman and
Eanna Flanagan) and Numerical Relativity (Richard Price and Carsten Gundlach).
The third was Mathematical Relativity (Rick Shoen, Hubert Bray).
I was very pleased with the degree to which all speakers gave talks
that were accessible to broad audiences.  As a result, the talks on line
at the above URL provide an excellent place for non-specialists to get
insight into topics from the recent proofs of the Penrose Inequalities to
the status of black hole collision simulations, and from
gravitational wave sources to microscopic
black hole entropy from both the loop and string points of view.
Rather than go into details here, I would strongly urge the reader to
go directly to the conference URL above, view the transparencies, and listen
to the talks and ensuing discussions.  In a few places the audio recordings
are a bit rough (due to interference from the nearby airport), but
there is no doubt that they will be a useful resource.

The conference webpage is 
\htmladdnormallink{http://www.itp.ucsb.edu/online/gravity\_c99/}
{http://www.itp.ucsb.edu/online/gravity\_c99/}

\newpage

\section*{\centerline {
Yukawa International Seminar}}
\addtocontents{toc}{\protect\smallskip}
\addcontentsline{toc}{subsubsection}{\it  
Yukawa International Seminar, by John Friedman}
\begin{center}
John Friedman, University of Wisconsin,  Milwaukee\\ 
\htmladdnormallink{friedman@thales.phys.uwm.edu}
{mailto:friedman@thales.phys.uwm.edu}
\end{center}
\parindent=0pt
\parskip=5pt

About 150 people gathered under cloudy skies in the old imperial
capital, Kyoto, June 28 - July 2, to listen to six days of talks and
view an extraordinarily good set of posters, all summarizing
theoretical, observational, and experimental work on gravitational
waves, black-hole physics, and numerical gravity during the Yukawa
International Symposium (YKIS99) on black holes and gravitational
waves.  Unfortunately, a summary of posters would make this already
long review too long to be useful.\\

In a first session on black holes in a quantum context, Ted Jacobson
summarized his work with Corley and Mattingly (e.g., hep-th/9908099).  
In the usual semiclassical computation of Hawking radiation, the
late-time flux arises from modes that, prior to the black hole's
formation, are vastly shorter than the Planck length; and one can worry
that black-hole radiance would not survive if the universe's
small-scale structure does not allow one to make sense of such
ultra-high frequency fields.  Jacobson and his coworkers address the
problem by placing a lattice on a black-hole background, and looking at
field theory on the lattice -- considering, in effect, quantum field
theory on a discrete spacetime.  Satisfyingly, the lattice reproduces
the Hawking effect with an accuracy that depends on the ratio of the
lattice spacing to the black hole radius.  A lattice regular at the
horizon is not static, and the lattice used is falling inward.  The
scattering by the lattice of a ingoing to outgoing wave is formally
analogous to the Bloch oscillation of an electron in in a crystal with
a uniform electric field.\\

Following this discussion, Gary Gibbons spoke on black holes in unified
theories. He noted that classical solutions are important in quantum
theories if their quantum corrections vanish, and that commonly
requires supersymmetry to cancel the fluctuations of bosons against
those of fermions.  The Breckenridge-Myers-Peet-Vafa solution provides
an example of a BPS black hole with nonvanishing angular momentum, a
solution that was used to count states using D-brane techniques (and
gave agreement with black-hole entropy).  Gibbons and Herdeiro
(hep-th/9906098) have completed a substantial study of the solution,
finding its geodesics and computing the scattering of a scalar field
off this extreme black hole. The solution includes an example of a ``naked
stable time machine,'' with spatially unbound geodesics that can travel
back in time; but Gibbons and Herdeiro argue that chronology protection
may be enforced by the third law of thermodynamics, preventing the
formation of an extreme black hole by means of a finite process. \\

Describing the work led by Israel's group on the nature of the 
interiors of black holes formed in collapse, Patrick Brady reviewed
the linear instability of the Cauchy horizon and the spherical 
models of black-hole interiors and then turned to more recent work 
that appears to confirm key features of the earlier models.  
In particular, Ori and Flanagan have used the Cauchy-Kovalevsky theorem to 
show that ``there exist functionally generic solutions of Einstein's
equations containing a null and weak scalar curvature singularity,''
and work by Barack and Ori and by Israel, Brady, Chambers, Droz and 
Morsink characterizes more precisely the Weyl curvature near these null 
Cauchy-horizon singularities.  \\

Andrzej Krolak continued the discussion of the nature of singularities
in gravitational collapse, summarizing theorems that characterize 
Cauchy horizons or restrict the
occurence of naked singularities. Here are a few.\\
Chrusciel and Galloway and Budzynski, Kondracki, and Krolak have shown
the existence of a large class of nowhere differentiable Cauchy
horizons.  Harada, Iguchi, and Nakao showed that generic
counterrotation prevents central shell-focusing formation. This is
consistent with Rendall's result with cylindrical symmetry that a
regular distribution function in phase space prevents naked
singularities of the kind apparently seen by Shapiro and Teukolsky,
using a singular distribution function for collisionless matter; and
consistent with the conjecture that matter described by smooth
distribution functions obeys cosmic censorship -- that, as in the
Newtonian theory,  velocity dispersion dissolves naked singularities.\\
              
Matt Choptuik summarized work by  about 30 people on critical phenomena 
in gravitational collapse that has led to a coherent picture.  Critical solutions are unstable by construction, lying on the boundary between two distinct stable endstates of collapse -- black hole or no black hole.  
Underlying the key features of near-critical collapse is the fact that 
the critical solutions are ``minimally unstable intermediate  attractors,'' solutions whose linear perturbations have a single unstable mode.  
Critical solutions exhibit discrete self-similarity (an oscillation 
within a scaling envelope) characterized by a rescaling exponent, 
for massless scalar, gravitational, and SU(2) Yang-Mills fields; 
while perfect fluids and multiple-scalar field systems are continuously
self-similar.  The transition to collapse studied earlier, of, say 
neutron stars pushed over their upper mass limit by an addition of 
an arbitrarily small mass, exhibits a mass gap, and Choptuik calls
these type I transitions. Examples are massive scalar fields, 
and colored black holes (variants with horizons of the Bartnik-Mckinnon 
Y-M Einstein solutions); the latter have overlapping regions in 
parameter space that correspond to type I and type II solutions. 
Choptuik's transparencies, with names and details suppressed here are at   
\htmladdnormallink{http://laplace.physics.ubc.ca/People/matt/Doc/ykis99.ps}
{http://laplace.physics.ubc.ca/People/matt/Doc/ykis99.ps}.

Jeffery Winicour described the characteristic treatment of black holes,
and the current status of the The Pitt null code, developed by Welling,
Isaacson, Gomez, Papadopoulos, Lehner, Bishop, Maharaj, Szylagyi, and
Husa.  In the vacuum case, a 3-D code has been implemented and tested
in a variety of contexts.  More recent work (
\htmladdnormallink{gr-qc/9901056}{http://xxx.lanl.gov/abs/gr-qc/9901056})
 incorporates
a perfect fluid with a 1-parameter equation of state.  It has so far
passed tests involving localized distributions of matter around a
Schwarzschild black hole, and the code is found to be stable and
convergent. Modifications are needed to handle shocks, and problems of
astrophysical interest remain to be tackled.  \\

Next morning, Kip Thorne began his talk by prodding the theorists to
intensify their effort to keep up with the rapid progress of the LIGO
experimentalists, and provide an accurate template for inspiral, with
no drift in phase over the time of detection.  Theorists particularly
lag in the NS/BH problem, where spin-induced precession is important.
The construction of the two LIGO observatories is essentially complete,
and LIGO I sensitivity is to be reached by November, 2001.  A LISA date
of 2008-2010 is likely (i.e., US support is likely).  Thorne emphasized
that already for LIGO II with signal recycling, the standard quantum
limit may be reached. To reach greater sensitivity, one cannot rely on
standard position detectors that ignore correlations, and Thorne
reviewed work with Braginsky, Gorodetsky, Matsko, Vyatchanin, Khalili,
Levin, and Kimble on measurements beyond the SQL.  Methods rely on
correlations between the photon shot noise and the back-action noise
induced by radiation pressure on the test masses. One can modify the
input or output optics of current interferometers so that measurements
at different times commute and state reduction has no influence on the
noise.  Thorne's current estimates for LIGO II event rates:  \\ NS/NS,
a few/yr; NS-BH a few/month (!);  BH-BH, unknown.\\ Other sources
mentioned were r-modes, strongly accreting LMXB's, and accretion induced
collapse of white dwarfs.\\

Seiji Kawamura spoke on the current status of the Japanese detector,
TAMA300:  a mode cleaner was locked successfully 4 months ago, and a
noise spectrum has been obtained.  He summarized the recycling
arrangement, and ongoing efforts to reduce shot, seismic and thermal
noise.  Great luck would be needed if there were to be a source strong
enough for TAMA300 to detect, but plans are to step up to a much 
larger, 3-km cryogenic detector in the Kamgoka tunnel. \\
 
Next, two talks,  by Ed Seidel and Jorge Pullin, on colliding black
holes.  Seidel noted both the long distance to go before inspiral can
be computed and the progress made by the NSF black-hole grand challenge
project, in developing a code that handles a variety of initial data
sets, giving, in particular a stable 3-D evolution of a set of
distorted black holes.  The development of the cactus code was
outlined, and a general PDE solver, due for public release in August,
with an open source code, may be of wide interest to readers of MOG.
10 groups are now using it. Seidel emphasized recent methods (initially
due to Shibata and Nakamura) of that give significantly more stable
evolution, by promoting badly behaved quantities to independent
variables.  Although evolutions can be very accurate, by $t = 50 M$,
every code crashes. \\

Pullin described work by a number of people on colliding black holes in
the close limit (
\htmladdnormallink{gr-qc/9905081}{http://xxx.lanl.gov/abs/gr-qc/9905081}).  
From a computational viewpoint,
binary black hole coalescence has three stages: a post-Newtian
inspiral, of about $10^4$ orbits, ending at roughly 10-12 $M$; the
plunge; and the ringdown.  The key to a perturbative treatment of the
ringdown stage is that realization that it is {\em is} the ringdown
stage, that a common, distorted horizon surrounds what were two black
holes well before their apparent horizons merge. The evolution of the
distorted horizon is what Pullin and his collaborators have treated
with remarkable success as a perturbation of a Schwarzschild black
hole. The work has included the development of a second-order 
perturbation formalism and its application to the ringdown problem.  
The approach serves as a code check for numerical relativity and
a way to allow dying codes to run longer; and the ringdown serves in
its own right as a source for LIGO.  An analogous close limit for
neutron stars, with the merged stars regarded as a perturbation 
of a spherical star has also been recently developed (
\htmladdnormallink{gr-qc/9903100}{http://xxx.lanl.gov/abs/gr-qc/9903100}).

Yasufumi Kojima and I then gave two talks on the recently discovered
r-mode variant of the nonaxisymmetric instability that besets rapidly
rotating relativistic stars.  First noticed by Nils Andersson, in a
numerical study, the instability of these axial-parity modes may
dominate the spin-down of neutron stars that are rapidly rotating at
birth, and the gravitational waves they produce may be detectable by
LIGO II with narrow banding from sources out to somewhere between 4 and
20 Mpc.  These dramatic implications have led to papers by about 35
authors in the past two years, but because my talk is now on the Web in
a version written with Keith Lockitch (
\htmladdnormallink{gr-qc/9808083}{http://xxx.lanl.gov/abs/gr-qc/9808083}), 
and each of our
talks at the ITP can be seen and heard live at
\htmladdnormallink
{http://www.itp.ucsb.edu/online/gravity\_c99/} 
{http://www.itp.ucsb.edu/online/gravity\_c99/}, 
I'll leave it at that.
Caveats are the assumption that nonlinear effects will allow the mode
to grow until perturbed velocities are of order of the background
velocity -- that the mode does not transfer its energy to turbulence or
to a magnetic field, while its amplitude is small. \\

Kojima emphasized research he has done, partly with Hosonuma, on the
r-modes of of relativistic stars, reporting work that showed a
continuous spectrum for nonisentropic stars in a slow-rotation
approximation, and Beyer and Kokkotas make the claim
precise. In addition, Kojima and Hosonuma have studied the
mixing of axial and polar perturbations to order $\Omega^2$ in rotating
relativistic stars, again finding a continuous mode spectrum.  Kojima
obtained a single, second-order equation for the radial behavior of the
modes.  If the continuous spectrum is a genuine feature of relativistic
stars, it would be remarkable, but it may well be an artifact of
approximations that force the frequency to be real: In the slow-motion
approximation, the continuous spectrum arises from the vanishing of the
highest derivative term in the ODE, found by Kojima, that describes the
axial-parity modes.\\

A session devoted to the post-Newtonian computations reported 
progress on calculations 
that must provide the highly accurate templates for binary 
inspiral that are needed to use gravitational-wave detectors for astronomy. 
Both Cliff Will and Luc Blanchet spoke on the post-Newtonian description
of binary inspiral.  Luc Blanchet spoke on the post-Newtonian description of
gravitational radiation, developed by Damour and colleagues.  He
emphasized the use of the Hadamard expansion to regularize the
infinities that arise at the position of point-particles, when one
expresses the multipole moments of the source as integrals extending
over the distribution of stress-energy.  One's confidence in using this
renormalization method relies on a combination of its success and its
elegance.\\

Cliff Will summarized a method called DIRE (Direct Integration of the
Relaxed Einstein Equations), based on a framework developed by Epstein
and Wagoner and extended by Will, Wiseman and Pati.  Like the
Blanchet-Damour-Iyer approach, DIRE begins with integrals over source
and field. The integrals are finite when one restricts the use of the
slow-motion approximation to the near zone and observes that the
far-zone integral is bounded for a source that is well-behaved in the
distant past.  Equations up to 3.5 PN order are obtained, within the
assumption that the orbiting bodies are sufficiently small, by
isolating terms that neither vanish nor blow up as the size $D$ of a
body shrinks to zero, with the remaining terms absorbed into
renormalized masses of the bodies.  The resulting procedure is
well-defined, although the assumption has been checked completely only
at 1PN order.  \\

In an enthusiastic update on the prospects for GEO600, Bernard Schutz
emphasized the expected performance with signal recycling in both a
narrow-band and broad-band mode: a maximum sensitivity of 
$h < 10^{-22}$ at minimum noise, for frequencies between 100 and 1000 Hz.
The state of the project: The vacuum system is complete, and the first
mode cleaner is locked and working; interferometry and test optics are
expected to be ready by mid-2000, and full sensitivity is to be reached
by mid-2001.  There has been close collaboration with LIGO, and a
memorandum of understanding has been signed for full data exchange
between LIGO I and GEO600.\\

The next day saw two talks on black-hole astrophysics.  Nils Andersson
summarized work on oscillations of rotating black holes, mentioning his
work with Krivan, Laguna, and Papadopoulos on a 2-D code that evolves
perturbations on a Kerr background and produces waveforms for
black-hole ringing.  Andersson emphasized two regimes of outgoing-mode
ringing for a given value of $l$, say $l=2$, corresponding to the
different imaginary parts of the frequencies for $m = \pm l$.  It
remiains to be seen whether, in the extreme Kerr limit both the
retrograde, more quickly dying mode and the prograde, slowly dying mode
both have comparable amplitudes (Mashoon and Ferrari suggested that the
prograde mode was suppressed).  Preliminary work suggests a curious
feature of near-extreme Kerr: a sum over harmonics appears to give an
oscillation whose damping is described by an envelope with a power-law
fall-off that is {\em slower} than the Price tail.  If true, one might
never see the Price power law (for $m\neq 0$) in the late-time ringing
of a near-extreme black hole. \\

Shin Mineshige spoke on the dramatic success of the ADAF
(advection-dominated accretion flow) model of accretion disks that 
started with the 1977 work of Ichimaru.  In this model, heat is
dominantly transported by radial gas motion, and the spectrum is
broad-band.  Mineshige emphasized the fact that three models of
accretion -- the standard thin and thick disk models and the ADAF model
are all solutions to the same set of equations with different values of
optical depth.  He reiterated the evidence for a central black hole
in ADAF disks, arising from the fact that the emitted power is much
smaller (as a fraction of the mass accretion rate) than it should be if
matter were falling on a solid surface. And he discussed the observational
tests of disk models and a recent model for X-ray novae. \\

Interspersed with these talks, and continuing through the next
morning's sessions were a series of talks on the binary coalescence
problem for neutron stars.  Eriguchi and Gourgoulhon began these with a
discussion of numerical work on Darwin Riemann problems in Newtonian
gravity and in GR.  The classical Roche, Darwin, and Riemann problems
refer respectively to (Roche) the tidal forces on a finite mass
orbiting a point mass with its spin and orbital frequencies identical;
(Darwin) both masses are finite perfect fluids; (Riemann) the masses
can have internal vorticity.  The Darwin-Riemann problem that Eriguchi
and Gourgoulhon consider has two masses whose spin frequencies are
arbitrary, but for which the planes of rotation are aligned with the
orbital plane.  A series of papers by Uryu and Eriguchi have
numerically solved the exact problem for Newtonian polytropes; and
Usui, Eriguchi, and Uryu have begun a program to construct
quasi-equilibrium models in GR, starting with spacetimes having an
exact Killing vector of the form $t^\alpha + \Omega\phi^\alpha$, using
a truncated set of field equations that allows a non-radiative field
and a smaller set of potentials than one would need for the exact
binary system.  They thus replace the approximation of spatial conformal flatness by one in which the metric has only a $t-\phi$ off-diagonal term.
Bonazzola, Gourgoulhon, and Marck repeat the Mathews-Maronetti-Wilson
computation for an $n=1$ polytrope (adiabatic index $\gamma = 2$),
using a multi-spectral method and obtaining agreement with the new MMW
code to better than 2\%.  No innermost stable circular orbit is found
for this value of $\gamma$.  A useful summary of our knowledge of
the ISCO was given (see 
\htmladdnormallink{gr-qc/9904040}{http://xxx.lanl.gov/abs/gr-qc/9904040} 
and Uryu-Eriguchi). 

Wai-Mo Suen and Ken-ichi Oohara summarized progress on the numerical
simulation of coalescing neutron stars by the NASA Neutron Star
Grand-Challenge group and by the Japanese group.  Suen's discussion
emphasized the extensive code testing that is underway and successes in
meeting the Grand-Challenge milestones.  A long enough time evolution
to model coalescence is still in the future, but Suen reported a
computation of head-on collisions using the coupled Einstein and
hydrodynamic equations. These runs tested a conjecture of Shapiro that
the shock-heating generated by infall, at least for stars falling from
infinity, is enough to support the star until neutino cooling sets in
(a time long compared to the dynamical timescale).  If true, this would
give an early cutoff to gravitational-wave emission. However, Suen
argued that the dynamical time scale of infall was so short that the
shock heating effect might not be important.  He showed a simulation of
the head-on collision of two 1.4 solar mass neutron stars, and reported
finding an apparent horizon in the infalling time scale.  Oohara
summarized the longer history of the Japanese program, with a full GR
code completed in '94.  Test runs of on the order of one revolution
have been run on a 201$^3$ grid, with a 10-hr CPU time on a VPP300.
Finally, Masaru Shibata has completed and tested on sample problems a
related 3-D code for binary coalescence of neutron stars
(\htmladdnormallink{gr-qc/9908027}{http://xxx.lanl.gov/abs/gr-qc/9908027}). 

Following the binary-coalescence series, Max Ruffert and Peter Mezaros
presented different views on the possible origins of $\gamma$-ray
bursts.  BATSE has revealed about one burst/galaxy/$10^6$ years,
distributed isotropically at distances of order $10^{28}-10^{29}$ cm.,
implying luminosities of order $10^53 erg/s$.  The duration of bursts
varies greatly, ranging from milliseconds to hours.  Evidence for a
fireball as a common source is good, but fireballs may be produced by
mergers of NS-NS, NS-BH, WD-NS, WD-BH, or by a collapsar, a rotating,
collapsing ``failed'' supernova (or possibly a neutron star pushed over
its upper mass limit by accretion).  Evidence for the collapsar comes
from the identification of $\gamma$-ray bursts with galaxies that
suggest the bursts probably occur in star-forming regions more often
than would be expected for the old NS-NS or NS-BH systems.  Ruffert and
his collaborators (Janka, Eberl, and Fryer, astro-ph/9908290) have run
a series of Newtonian simulations of NS-BH and NS-NS mergers
incorporating back-reaction of gravitational waves.  Using a
Lattimer-Swesty equation of state and carefully taking account of
neutrino sources and sinks, they confirm BH-NS mergers as a possible
source of $\gamma$-ray bursts, but find an energy of 10$^51$ erg
requiring Lorentz beaming at the upper end of the possible.  Mezaros'
even-handed summary reviewed evidence for the fireball model he
developed with Rees and others.  Fireballs from all of the mechanisms
have similar energies, with $10^{54}$ erg possible via MHD, while less
than $10^{erg}$ is likely if only neutrino annihilation is used (as in
the simulations discussed by Ruffert). This leaves coalescence clearly
still in the game.\\

The final two talks, by Bernard Carr and Jun'ichi Yokoyama, concerned
primordial black holes.  Carr delineated ways that PBH's can be used as
a probe of the early universe; in particular, the limit on PBH's set by
the absence of observed evaporating black-holes limits the spectral
index during inflation.  Both Carr and Yokoyama emphasized the
possibility that MACHOS are PBH's of mass 0.5 $M_\odot$, the right mass
for their having been created in a quark-hadron transition. Should the
LMC MACHOS be PBH's, work by Yokoyama and collaborators sharply
constrains  parameters of inflationary models; and nearby BH-BH
coalescence would be frequent, with possibly observable gravitational
waves (Nakamura, Sasaki, Tanaka, Thorne).  Carr considered a
PBH-related test of whether $G$ varies.  
   
\newpage

\section*{\centerline {
Minnowbrook symposium on the structure of space-time}}
\addtocontents{toc}{\protect\smallskip}
\addcontentsline{toc}{subsubsection}{\it  
Minnowbrook symposium on the structure of space-time, by Kamesh Wali}
\begin{center}
Kameshwar Wali, Syracuse University\\
\htmladdnormallink{wali@suhep.phy.syr.edu}
{mailto:wali@suhep.phy.syr.edu}
\end{center}
\parindent=0pt
\parskip=5pt

This century, with Einstein's general theory of relativity, space-time
assumed a dynamical role in the theory of gravitation. It was a
revolution in our ideas and has led to momentous discoveries regarding
the large scale organization of matter in the universe. Equally
important is the progress we have made in exploring the subatomic
world and in the search for an all-unifying theory of fundamental
interactions. But problems remain and physics is at cross roads of
varying ideas at the "End of the Millenium."

The Minnowbrook Symposium took place on May 28-31 and was organized to
bring together the exponents of varying ideas, specifically,
1). Classical general relativity, 2). Recent developments in string
theory and the emerging view of space-time, 3). Quantum gravity and
generalized quantum field theories and, 4). Non-commutative geometry
and its perspective on the structure of space-time at short distances.

The first day of the symposium in the morning began with John
Stachel's review of the various space-time structures associated with
no-relativistic Galileian, and relativistic Minkowskian theories, and
showing how generally covariant space-times are fundamentally
different from their predecessors and the implication of this
difference on the usual starting point of space-time theories. His
talk was followed by two talks on Black Hole Thermodynamics and
related problems. Robert Wald, after reviewing some of the established
results in black hole thermodynamics, spoke about the major unresolved
issues such as whether black hole entropy should be viewed as
"residing" in its deep interior, on its horizon, or in its "thermal
atmosphere." Ted Jacobson gave an overview of some results and open
questions concerning the meaning of black hole entropy and the nature
of the holographic bound on the entropy contained within a surface of
given area. He dealt with issues such as the role of internal states,
entanglement, species independence, renormalization of G and the
Second Law of Thermodynamics. He ended up providing a derivation of
the Einstein's equation from the proportionality of entropy and
area. In the afternoon, Ted Newman spoke about his (and his
collaborators) alternate formulation of general relativity in terms of
families of characteristic surfaces and a scalar function. With these
as fundamental variables, the conventional variables such as the
metric tensor become derived concepts and the final equations,
although they do not resemble the standard version of GR, yield
results identical to those of GR. He was followed by Roger Penrose,
who expounded his ideas regarding quantum state-vector reduction,
viewing it as a gravitational phenomenon. He discussed both
theoretical arguments as well as possible experimental tests (now
actively pursued) and presented some new theoretical developments.

The first day ended with a lively panel discussion led by Lee
Smolin. Different subgroups got together first to discuss important
questions in light of the day's proceedings and issues to be addressed
the following days.

The second day began with a session on String Theory and String
Theorists' view of Space-time. Due to illness, Joe Polchinski had to
cancel his participation at the last minute, but we had two excellent
talks by Michael Douglas and Brian Greene that covered some aspects of
Polchinski's subject matter. Speaking about D-Geometry, Matrix theory
and Noncommutative geometry, Douglas surveyed recent developments in
D-branes theory and the nature of space-time as seen by
D-branes. Greene presented new geometrical ideas that have emerged out
of recent researches in String Theory, ideas involving dualities,
mirror symmetry, topology change, and non-commutative geometry, giving
rise to an evolving "quantum geometry" or "string geometry." In the
afternoon session, Abhay Ashtekar discussed the features of quantum
geometry through two specific examples, namely, Einstein-Maxwell
theory in 2+1 dimensions and quantum geometry of black hole horizons
in 3+1 dimensions. With non-perturbative, complete solution of the
problem, one finds unexpected limitations of the classical and
semi-classical theory in the first case. In the second case, the
horizon geometry is described by the quantum Chern-Simons theory on a
punctured 2-sphere, giving rise to states that account for the black
hole entropy. John Baez explained how spin network techniques have
provided a mathematically rigorous and intuitively compelling picture
of the kinematical aspects of loop quantum gravity. For a true
understanding of the dynamical aspects of gravity, he said, one needs
a model for 'quantum 4-geometry", that is a truly quantum mechanical
description of the geometry of space-time. He discussed the notion of
" spin foam", as a probable candidate for such a description.

The last and final day began with Alain Connes who described the
foundations of non-commutative geometry in terms of spectral triplets
and how it has been successful in predicting some features of the
standard model. He outlined a "Spectral Action Principle" from which
one could derive elementary particle interactions as fluctuations of
he metric. Ali Chamseddine took it from there and discussed in some
detail examples of non-commutative spaces appearing in the standard
model and string theory. He showed how when the spectral action
principle is applied to the standard model, internal symmetries merge
with space-time symmetries, unifying gravity with the other
interactions. Dirk Kreimer discussed recent developments in
perturbative quantum field theories pertaining to the role of a Hopf
algebra governing their renormalization and how this algebra is
related to the Hopf algebra structure of diffemorphisms in the context
of non-commutative geometry. The afternoon session was devoted to
Discrete as opposed to a continuum space-time picture at short
distances. Klaus Fredenhagen spoke about fundamental uncertainty
relations for space-time coordinates and how they can be modeled in
the framework of non-commutative geometry. Rafael Sorkin described " A
Causal Set Dynamics," showing how, starting from certain causality
conditions and a discrete form of general covariance, one can derive a
very general family of classically stochastic, sequential growth
dynamics for causal sets. The resulting theories, according to Sorkin
, provide a relatively accessible "half-way house" to full quantum
gravity. He noted that non-gravitating matter can also arise
dynamically in such theories. The final talk of the session was due to
David Finkelstein on Spin, Statistics and Space-time Structure. His
basic idea was that below the quark scale, matter, measurement, and
space-time can no longer be separated operationally. They fuse into
one variable.
 
The final session organized by Carlo Rovelli was devoted to short
presentations on some important topics that were not covered in the
other sessions. These included among others, Amanda Peet who spoke
briefly on dualities in string theories and their implications, Julius
Wess on some recent developments in Quantum Groups and Roger Penrose
on Twistors.

From all accounts, the symposium achieved its main purpose in that it
provided an opportunity for the presentation of varying points and
in-depth discussions on fundamental questions regarding space-time
structure. Minnowbrook Conference center with its idyllic
surroundings, its seclusion and comfort generated a stimulating and
friendly atmosphere among the forty or so participants.

For those interested, the complete information regarding the symposium
is located at the following URL:

\htmladdnormallink{http://www.phy.syr.edu/research/he\_{}theory/minnowbrook/}
{http://www.phy.syr.edu/research/he\_{}theory/minnowbrook/}

If you click on the speaker's name in the program section, you can
link to his transparencies (if the click does not work, add the "
speaker name.html" (i.e. Penrose.html) to the /minnowbrook link given
above.

\newpage

\section*{\centerline {
Black holes II and CCGRRA 8}}
\addtocontents{toc}{\protect\smallskip}
\addcontentsline{toc}{subsubsection}{\it  
Black holes II and CCGRRA 8 by Jack Gegenberg and Gabor Kunstatter}
\begin{center}
Jack Gegenberg, Univ. of New Brunswick and Gabor Kunstatter,
Univ. of Winnipeg\\
\htmladdnormallink{lenin@math.unb.ca}
{mailto:lenin@math.unb.ca}
\htmladdnormallink{gabor@theory.uwinnipeg.ca}
{mailto:gabor@theory.uwinnipeg.ca}
\end{center}
\parindent=0pt
\parskip=5pt

Following a tradition established two years ago, the Black Holes
Workshop and the Canadian General Relativity and Relativistic
Astrophysics Conference were held consecutively and in nearby locales.
Black Holes II: Theory and Mathematical Aspects, took place June
6-9,1999 at the resort town of Val Morin in the Laurentian Mountains
north of Montreal, Quebec, followed closely by CCGRRA 8 (June 10-12)
at McGill University in Montreal.  Both conferences were supported by
the Canadian Institute of Theoretical Astrophysics, the Centre de
Recherches Mathematiques.  Black Holes II also received financial
support from the Canadian Institute for Advanced Research.  The local
organizers for both meetings were Cliff Burgess and Rob Myers. The
organizers must be congratulated for putting together two very
successful and exciting complementary meetings.
\begin{center}
Black Holes II
\end{center}
The talks and discussions in Black Holes II revolved around the nature
of and origins of black hole thermodynamics, with discussion often
considerably heated by the dialogue between string theorists and
relativists.  The morning sesssions consisted of three plenary talks
and were followed by a sumptious lunch. In the afternoon the
contributed talks were presented, and after an excellent dinner, the
final plenary session of the day took place.  Discussions often
continued long into the night in the resort's bar. These discussions
sometimes came to definitive and startling conclusions that were
rarely remembered by morning.

The first speaker at the workshop was Bob Wald, who set the stage by
reviewing the geometric basis of black hole thermodynamics.  Abhay
Ashtekar continued in this spirit by describing his concept of
isolated horizons, and the associated quasi-local definitions of mass
and surface gravity.  The morning session concluded with Werner
Israel's re-appraisal of `t Hooft's brick wall model of black holes.

After interesting contributed talks by Renaud Parentini, Claude
Barrabes, Bei-lok Hu and Alessandro Fabbri, Leonard Susskind spoke
during the first evening session on how to understand superluminosity
in the context of the holographic conjecture.

On the second day, Steve Carlip spoke on his construction of black
hole statistical mechanics from states induced on the boundary/event
horizon of black holes in spacetimes of arbitrary dimension.  Gabor
Kunstatter presented a general discussion of 2D black hole
thermodynamics followed by a summary of recent calculations of quantum
corrections to the thermodynamic properties of charged 2-D black
holes.  Finally Emil Martinec discussed the interplay of D-branes and
black hole thermodynamics in the M-theory context. The afternoon
session consisted of contributed talks by Amanda Peet, David Kastor
and Daniel Kabat, Jennie Traschen and Simon Ross.  The plenary talk in
the evening was given by Don Page, who described the thermodynamics of
nearly extreme black hole under various assumptions about the
degeneracy of the ground state and the density of nearby states.

On the last full day of the workshop, Ted Jacobson forcefully
presented his perspective on the question of the `ultravioletness' of
the source of Hawking radiation, provoking lots of discussion from the
mixed stringy/ relativist participants.  Valeri Frolov described his
program for understanding the origin of black hole thermodynamics from
the statistical mechanics of the constituent matter fields in `induced
gravity'.  Robert Mann then outlined a proceedure for determining the
boundary terms in black hole thermodynamics inspired by the AdS/CFT
conjecture. That afternoon, the contributed talks were given by Julian
Lee, Ivan Booth, Jack Gegenberg, Roberto Casadio and Martin Rainer.
The evening talk by Bill Unruh presented the case for taking seriously
the sonic analogue of black holes (`dumbholes').  This too sparked
considerable debate between those who believed that black hole
thermodynamics contained important microscopic properties that were
missing in the dumbhole analogy, and those who felt that all the
essential (infrared) features were adequately represented in the
model.

On the last (half) day of the workshop the only plenary talk was
delivered by Juan Maldecena. This talk entailed a discussion on some
issues in the AdS/CFT conjecture, in particular on peculiar effects in
the boundary CFT. Th e conference closed with three interesting
contributed talks, by Finn Larsen, David Lowe and Steve Gubser.

As the participants boarded boarded the bus for Montreal after the
last talk, many felt that the controversy pointed to an overall
unsettled state of affairs in fundamental physics and signalled
interesting times ahead...
\begin{center}  CCGRRA 8
\end{center}
Many of the Black Hole II participants went from Val Morin directly to
Montreal for CCGRRA 8, where they were joined by many others from
across the spectrum of gravitational physics.  CCGRRA 8 participants
were able to take advantage of Montreal's many fine restaurants and
even had the rare opportunity of watching the Expos win a baseball
game!

On the first day, we heard from Peter Saulson about progress in
building gravity wave detectors. This was followed by Vicky Kaspi's
talk on the state of binary radio pulsar timing as a GR effect and
then by Jeff Winnicour's discussion of numerically calculating the
dynamics of black hole collisions.  After lunch, Sharon Morsink
demonstrated that certain instabilites in rotating neutron stars may
explain the properties of some recently discovered pulsars.

On the second day, Ted Jacobson discussed open questions about black hole
entropy, in particular those arising from a conjectured holographic bound.
Abhay Ashtekar continued the discussion he begin at Black Holes II, here
using quantum geometry to describe black hole thermodynamics, and
comparing the result to the stringy discussion.  The morning session
concluded with an introductory talk by Leonard Susskind on the
holographic principle, emphasizing its origins in black hole quantum
theory and string theory.   Finally, Gilles Fontaine built the case for
white dwarfs as a major component of dark matter.

On the last day, we first heard from Lev Kofman on how to build better
inflationary models incorportating preheating.  This was followed by
Bill Unruh's discussion of second order perturbations in the expansion of
the universe.  The final talk, by Don Page,  was on the
possibility of testing the
many-worlds viewpoint via quantum cosmology.

The contributed talks in the afternoon parallel sessions were in general
excellent.
Many of these talks were given by grad students and postdocs, and provided
an exciting introduction to the breadth and depth of gravitational
physics research today. It was clear to all the participants that this
 conference once again successfully fulfilled the CCGRRA's mandate to bring
 together
Canadian and international researchers in order to discuss the latest
 developments in Relativity and Astrophysics.

\newpage

\section*{\centerline{Hartlefest \& 15th Pacific Coast Gravity Meeting}}
\addtocontents{toc}{\protect\smallskip}
\addcontentsline{toc}{subsubsection}{\it  
Hartlefest \& 15th Pacific Coast Gravity Meeting by Simon Ross}
\begin{center}
Simon Ross, UCSB\\ 
\htmladdnormallink{sross@cosmic.physics.ucsb.edu}
{mailto:sross@cosmic.physics.ucsb.edu}
\end{center}
\parindent=0pt
\parskip=5pt

\begin{center}

\end{center}

A one-day meeting in celebration of Jim Hartle's 60th birthday was
held at the ITP at UCSB on the 25th of February. It was followed by
the 15th Pacific Coast Gravity Meeting, which took place on the 26th
and 27th. Hartlefest had a relaxed schedule, with four invited
speakers representing a few of the areas Jim has been interested
in. Most people attended both meetings, making this one of the larger
PCGMs, with a packed schedule of fifty-one speakers, and large
audiences.

At Hartlefest, Gary Horowitz led off with a discussion of the answers
string theory gives to some questions in quantum gravity, and went on
to review the recent conjecture by Maldacena relating gauge theory and
string theory. He was followed by Karel Kuchar, who supplemented his
talk on the quantization of diffeomorphism-invariant systems with some
anecdotes about Jim and some fashion tips. Kip Thorne talked about the
effect of tides on the calculation of gravitational waveforms from
binary systems, and about daring ideas to beat the standard quantum
limit in gravity wave detectors by not measuring the positions of the
test masses. The closing speaker was Murray Gell-Mann, who gave a
non-technical talk on generalized quantum mechanics and how the
familiar classical world could emerge as an approximate description.

Following the talks, there was a reception and then a buffet dinner at
the ITP. Once all this food and wine had worked their magic, some
brave people got up to honor Jim in a variety of entertaining
ways. Gary Horowitz and a group of assistants led the audience in
singing the relativity hymn (see
\htmladdnormallink
{http://cosmic.physics.ucsb.edu/hymn.html}
{http://cosmic.physics.ucsb.edu/hymn.html}
). Matt Fisher presented a
version of `The Cat in the Hat' written with his family in honor of the
contributions of Jim and others to the ITP. A large number of people
shared their memories of Jim, and everyone had a good time. The
scientific part of Hartlefest is on-line at
\htmladdnormallink{http://www.itp.ucsb.edu/online/hartle\_c99/}
{http://www.itp.ucsb.edu/online/hartle\_c99/}. To encourage a relaxed
atmosphere, the after-dinner remarks were not recorded.

The Pacific Coast Gravity Meetings are a popular regional
institution. They provide an opportunity to get together with
colleagues, catch up on gossip, and hopefully learn some interesting
new physics. The organization is very informal, to encourage
participation by the widest possible range of people. Anyone who wants
to speak can, and all speakers get the same amount of time.  In the
same spirit, I will attempt to briefly review all the sessions here,
although there is clearly a risk that no information will survive the
compression.

The meeting opened with a session of talks on cosmology. Daniel Suson,
Warner Miller and Kent Harrison focussed on classical aspects of the
early universe. David Salopek argued that Hamilton-Jacobi methods are
useful for both classical and quantum studies. Richard Woodard made
the striking proposal that the incorporation of quantum back-reaction
provides a natural mechanism for ending inflation. Later in the
meeting, Beverly Berger and Jim Isenberg presented results of
numerical studies of cosmological singularities.

With LIGO nearing completion, it was no surprise that there were a
number of talks related to production and detection of gravitational
waves. Scott Hughes, John Whelan and Lior Burko discussed different
approaches to calculating the effect of radiation reaction on the
sources. Richard Price explained the close limit in rotating black
hole mergers, in which a perturbed Kerr black hole is used as an
approximate description. In a later talk, Alcides Garat argued that
there is no conformally flat slicing of Kerr, which complicates the
study of these perturbations. William Krivan and Zeferino Andrade
discussed the tail in the waveform, which comes from quasi-normal
ringing of the resulting object. Patricia Purdue showed tidal effects
are gauge invariant, filling in some of the picture Kip Thorne had
sketched on the previous day. Lee Lindblom reviewed the r-mode
instability of rapidly rotating neutron stars, and discussed detection
prospects.

On the detection side, Teviet Creighton considered a surprising noise
source: the gravitational attraction between the test masses and
wind. Massimo Tinto addressed data analysis and unequal arm lengths in
a space-based interferometer. Jolien Creighton and Gabriela Gonzalez
discussed ways of dealing with noise in the detectors. Shane Larson
gave a talk on using gravity wave signals to limit the mass of the
graviton.

There were also talks on more general numerical work. Carsten Gundlach
and David Garfinkle used the self-similar critical solution to explain
scaling properties observed in numerical studies of gravitational
collapse. James Bardeen, James York and Luisa Buchman described the
program of constructing a hyperbolic system out of general
relativity. Buchman described some numerical results obtained by
implementing one of these systems.

The talks on quantum gravity covered a wide range of approaches and
issues. Herbert Hamber discussed using a homemade supercomputer
built from PCs to obtain results in a Regge theory approach. Jorge
Pullin gave a thorough review of the history and recent progress in
the quantum geometry program. Bryce DeWitt and Charles Torre described
some techniques in the superspace formulation. Alejandro Corichi
argued that reconstructing a physical geometry from a point in the
phase space may be non-trivial. Sharmanthie Fernando and David Kastor
talked about states where the metric must be an operator-valued
function. Jennie Traschen demonstrated agreement between a Born-Infeld
field theory and a spacetime picture of a string-threebrane
system. Kirill Krasnov argued that quantum gravity can be represented
as a constrained version of the topological BF theory, which might
provide an interesting formal approach to the path integral.

The effects of quantizing fields on a classical background continue to
be an active subject for research. Bill Hiscock re-examined the
disruption of the chronology horizon in Misner space. Paul Anderson
and Brett Taylor discussed backreaction effects in extremal and nearly
extremal black holes. Ted Jacobson discussed the effects of changing
the dispersion relation for the quantum fields.  Michele Vallisneri
attributed the Unruh effect to the failure of classical special
relativistic ideas in curved backgrounds.

Mathematical aspects of classical gravity were discussed at several
points in the meeting. Kristin Schleich proved a theorem relating
black hole topology to the topology of scri in asymptotically anti-de
Sitter spacetimes.  Don Witt talked about the uniqueness of locally
static solutions with a cosmological constant. Robert Mann showed that
in 1+1 gravity, the two-body problem can be exactly solved. Arthur
Fischer talked about general relativity as an unconstrained dynamical
system. Frank Estabrook and Andre Wehner offered classical
alternatives to general relativity motivated by symmetries. Micheal
Martin emphasized the similarity between the contracted Christoffel
symbols and a gauge field. Tevian Dray discussed the potential of
octonionic fermions to explain the features of the standard model.
William Pezzaglia proposed a new approach to spinning particles
involving polygeometric spaces. Leonard Abrams and Homer Ellis
proposed alternatives to the Schwarzschild solution.

At the end of the meeting, the Bell prize for the best student talk
was awarded to Teviet Creighton of Caltech for his talk on
"Atmospheric gravity gradients: a low-frequency noise limit for
LIGO". Some information about PCGM is available at:

\htmladdnormallink{http://cosmic.physics.ucsb.edu/pcgm.html}
{http://cosmic.physics.ucsb.edu/pcgm.html}.  

Next year's PCGM will be at
Los Alamos National Lab, organized by Warner Miller.

\newpage

\section*{\centerline {Second Capra workshop on Black Holes} \\
\centerline{and
Gravitational Waves}}
\addtocontents{toc}{\protect\smallskip}
\addcontentsline{toc}{subsubsection}{\it
Second Capra workshop by Patrick Brady and Alan Wiseman}
\begin{center}
Patrick R Brady, University of Wisconsin-Milwaukee\\
\mbox{\ }
\htmladdnormallink{patrick@gravity.phys.uwm.edu}
{mailto:patrick@gravity.phys.uwm.edu}
\\
Alan G Wiseman, University of Wisconsin-Milwaukee\\
\mbox{\
}\htmladdnormallink{agw@gravity.phys.uwm.edu}
{mailto:agw@gravity.phys.uwm.edu}
\end{center}
\parindent=0pt
\parskip=5pt

An informal workshop took place at University College Dublin, Ireland
from 16-20 August 1999.  The workshop emphasized theoretical methods
to investigate gravitational-wave generation, implications of
gravitational-wave astronomy for theorists, and identification of
important open problems.

There were 18 participants at the workshop.  During the first three
days, there were 12 talks by participants.  Each talk was 30-50
minutes in duration followed by equal time for discussion.  An outing
to nearby Powerscourt and Glendalough was organized for Thursday.
Friday morning was devoted to informal presentations by the remaining
participants.  The workshop was a huge success; it was generally
agreed that the format made for an excellent atmosphere to share and
discuss the many results that were presented.

Steve Detweiler made the first presentation.  His work (in
collaboration with Lee Brown) uses a Regge-Wheeler formalism to
compute the $O(\mu/M)$ corrections to geodesic motion for a particle
orbiting a Schwarzschild black hole.  He presented preliminary results
which indicate that the orbital frequency is reduced at the innermost
stable circular orbit when compared to the test particle limit.
Detweiler's talk set the tone for the workshop by reporting new
results from work in progress, and by encouraging open discussion of
the material.

Scott Hughes reported on progress computing the effects of radiation
reaction without the use of radiation reaction forces.  He showed
results for the radiative evolution of orbits and gravitational
waveforms generated by small objects in circular, non-equatorial
orbits around Kerr black holes.  Perhaps the most interesting result
is that the tilt angle of inclined orbits changes extremely slowly ---
more slowly in the strong-field than would be predicted by
post-Newtonian theory.

Kostas Glempedakis and Sathyaprakash reported on work in progress at
Cardiff. Kostas discussed ongoing work on radiation reaction effects.
Sathya discussed the $P$-approximants: a class of approximate
waveforms that model gravitational waves from inspiralling and
coalescing compact binaries.  The waveforms are constructed using two
inputs:
(a) two new energy-type and flux-type functions and (b) the systematic
use of Pad\'e approximation for constructing successive approximants
of these new functions.  The new $P$-approximants have larger overlaps
and smaller biases than the standard Taylor-series approximants. They
also converge faster and monotonically.  The presently available
$\left (v/c\right )^5$-accurate post-Newtonian results can be used to
construct $P$-approximate waveforms that provide overlaps with the
``exact'' waveform larger than $96.5\%,$ implying that more than 90\% of
potential events can be detected with the aid of $P$-approximants as
opposed to a mere 50\% that would be detectable using standard
post-Newtonian approximants.

Alan Wiseman discussed methods of computing the instantaneous radiation
reaction force on scalar charges orbiting Schwarzschild black holes.
These calculations utilize the DeWitt-Brehme method of regularization,
where the force is computed from the ``tail'' of the Green's function.
He also reported on progress computing the forces both from the mode
sum of the Green's function and analytic expressions for the Green's
function.

Amos Ori followed Wiseman by presenting a concrete a method of
computing the radiation reaction force with a mode sum.  Although the
contribution of each individual $l,m,\omega$ mode to the self-force is
finite, the naive sum of these contributions generically diverges (for
certain components of the force). Amos presented a new method for
regularizing the mode sum. In this method the mode-sum procedure is
modified in a suitable way, in order to obtain the correct (and
finite) expression for the self-force.

Lior Burko presented simple, though non-trivial, applications of Ori's
prescription for mode-sum regularization of self forces. For the cases
of static scalar or electric charges in Schwarzschild spacetime,
Burko's calculations recover the results known from completely
different procedures. Burko also presented results for the self force
acting on a scalar charge in circular motion around a Schwarzschild
black hole in the strong field regime. The conservative self-force is
attractive, and scales like the black hole's mass squared.

Warren Anderson discussed work in progress with \'{E}anna
Flanagan. They use dimensional and counting arguments to calculate the
contribution to the gravitational radiation reaction force from the
normal neighborhood of a massive particle in an arbitrary vacuum
spacetime.  Their method involves expansions in (geodetic
distance)/(radius of curvature) and (particle mass)/(radius of
curvature).

Brian Nolan spoke about two subjects. He reported on work on the
strength of the central singularity in spherical gravitational
collapse.  He also gave a brief critical review of attempts to define
black holes in arbitrary spacetimes using surfaces foliated by
marginally trapped 2-surfaces.

Eric Poisson reported on his own calculation of the Price power-law
decay of radiative fields (scalar, electromagnetic, and gravitational)
in Schwarzschild spacetime. His approach has the advantage of linking
the behavior of the field at future null infinity and near future
timelike infinity in a natural way: The two different power indices
arise by taking two different limits of a single expression. His
method also yields the field's behavior on the event horizon,
complete with all multiplicative factors.

\'{E}anna Flanagan talked about the non-detectability of squeezed
statistical properties of relic gravitational waves from the early
Universe, based on work with Allen and Papa
(\htmladdnormallink{gr-qc/9906054}
{http://xxx.lanl.gov/abs/gr-qc/9906054}).  He
reported, contrary to a claim of Grishchuk
(\htmladdnormallink{gr-qc/9810055}
{http://xxx.lanl.gov/abs/gr-qc/9810055}), that
even if relic gravitational waves are detected by LIGO/VIRGO or LISA,
their squeezed nature will not be detectable.  The squeezing is only
detectable by experiments whose duration is comparable to the age of
the Universe at the time of measurement.  The analysis does not rule
out indirect detections of the squeezing via measurements of the CMBR,
since CMBR photons measure the stochastic background over a time
comparable to the age of the Universe at recombination.

Nils Anderson presented recent results concerning the late-time
behavior of perturbations in the Kerr geometry. He discussed analytic
calculations based on the contribution from the black hole's virtually
undamped quasinormal modes that indicated that the field would
oscillate with an amplitude that falls of inversely with time (much
slower that the familiar power-law tails). The analytic work was
supported by numerical evolutions using the scalar field Teukolsky
code.  He suggested that this is a new (superradiance related)
phenomenon in black hole physics.

During a spectacular day in the Wicklow countryside on Thursday,
discussion continued on the many aspects of modeling gravitational
wave generation from astrophysical systems.  At the final session on
Friday morning, Dan Kennefick gave a short report on his recent work
studying the sociology of the community of theorists working on
gravitational waves.  The presentation was received with keen interest
by the audience, some of whom had participated in one of the issues
discussed by Dan.   Jolien Creighton discussed an approach to
identifying the gauge invariant component of the work done by tidal
forces in a binary system.   In a short presentation,  Patrick Brady
suggested that the Kerr-Schild form of black hole spacetimes might
be useful when computing approximate expressions for the radiation
reaction force on objects orbiting black holes.  Finally,  Ben Owen
discussed the relevance of gravitational radiation reaction for the
$r$-modes of nascent neutron stars.

In conclusion, Adrian Ottewill's excellent organization contributed
significantly to the success of the second Capra workshop on
gravitational waves.  The workshop was wonderfully paced and
organized, allowing a great degree of participation by everyone.  An
atmosphere of lively debate was facilitated by the allotment of equal
time for talks and discussion.

And yes, ``Capra'' does refer to Frank Capra, the American movie
director.  Frank Capra was a Caltech alumnus and benefactor.  The
first ``Capra'' workshop took place last summer at his ranch outside of
Los Angeles.

The workshop was supported in part by funds from University College
Dublin.  More information can be found at
\htmladdnormallink{http://www.lsc-group.phys.uwm.edu/\~{}patrick/ireland99/}
{http://www.lsc-group.phys.uwm.edu/\~{}patrick/ireland99/}.

\newpage
\section*{\centerline {
Third Edoardo Amaldi Conference on Gravitational Waves }}
\addtocontents{toc}{\protect\smallskip}
\addcontentsline{toc}{subsubsection}{\it
Third Edoardo Amaldi Conference, by Gabriela Gonz\'alez}
\begin{center}
Gabriela Gonz\'alez, Penn State\\
\htmladdnormallink{gig1@psu.edu}
{mailto:gig1@psu.edu}
\end{center}
\parindent=0pt
\parskip=5pt

The EDOARDO AMALDI CONFERENCE ON GRAVITATIONAL WAVES has been
designated as the cornerstone conference for the newly formed
Gravitational Wave International Community. The Amaldi Conference has been
held twice before, in Frascati, Italy, (1994) and at CERN, Geneva,
Switzerland, (1997), but takes on a new significance, now that it is
the main meeting for the Gravitational Wave International Community.

It  was held on the Caltech campus from July
12 - 16, 1999. About 300 participants, from all over the world
attended the conference. After the welcome remarks by Caltech
president. Dr. D. Baltimore, we heard a very moving and inspiring talk
by Ugo Amaldi, son of Edoardo Amaldi, on a life dedicated to broaden
the horizon of physics.

The first couple of days were dedicated to overview talks, while workshops
on more specialized topics were held later in the week. The conference did
not have parallel sessions, which meant that everybody learned much
more than the latest developments in their particular areas. This made
the conference more interesting than usual, even though the number of
talks was more limited.

The overview talks on the status of interferetry and bar detection of
gravitational waves (by Ruediger and Coccia, respectively), were a
preamble to the very exciting reports on the different projects around
the world, and even outside this world (LISA). Most of the
interferometers are doing much progress in their construction, while
LISA is arousing more interest in NASA. Bar detectors keep taking data
and improving their sensistivities getting close to the quantum
limit. We all expect that by the next Amaldi conference reports will
be on the cooperative enterprise of a world comunnity ready to learn
from fresh data from the interferometers.

On the astrophysical side, Kip Thorne (apart from giving a public
lecture) lead a discussion session with reports of different
scenarios for production of gravitational waves.

On the technical side of interferometers, there was a session with
very nice overview talks of the more important issues that limit the
detectors' sensitivities: suspensions, thermal noise, optical
configurations and quantum limitatins. Even the theorists who attended
these talks could follow them: we hope this is becomes a trend that
unites the community in their understanding and tackling of all these
difficult problems.

There were then workshops on each of those topics, lead by
P. Fritschel (configurations), E. Gustafson (lasers and optics), and
myself (suspensions and thermal noise). it was clear from these sessions
that the field is bubbling with new ideas which need people and time
to test, but undoubtedly will result in much improved interferometers. 

There was also a session on gravitational wave detection in space,
leaded by K. Danzmann, and a workshop on bar antennae, leaded by
W. Hamilton. These sessions proved that even such a new field as this
has already a history of detectors being pushed to their quantum
limits, and future projects to go in space!

The conference ended with a a session on signal processing and data
analysis, organized by B. Allen and S. Vitale. We heard about data
from satellite experiments, interferometer prototypes, bar detectors,
and we learned from the astronomers' experience with the Supernovae
Neutrino Network. 

We all thank S. Meshkov, the wonderful secretary, V. Kondrasho, and
the web page keeper, B. Kratochwill. The conference's web page is
still up, at 
\htmladdnormallink{http://131.215.125.172/}
{http://131.215.125.172/} , and can be consulted for
abstracts or more information on the program.

\newpage
\section*{\centerline {
Strings 99}}
\addtocontents{toc}{\protect\smallskip}
\addcontentsline{toc}{subsubsection}{\it
Strings 99, by Thomas Thiemann}
\begin{center}
Thomas Thiemann, Albert Einstein Institute, Golm, Germany\\
\htmladdnormallink{thiemann@aei-potsdam.mpg.de}
{mailto:thiemann@aei-potsdam.mpg.de}
\end{center}
\parindent=0pt
\parskip=5pt

This year's major conference on string theory, STRINGSS99, took place 
in Potsdam, Germany. The conference site was at the University of Potsdam 
close to the famous Sanssouci Park and the Neues Palais.
The meeting was mainily organized by the Albert-Einstein-Institut,
Max-Planck-Instititut f\"ur Gravitationsphysik, in particular 
by Olaf Lechtenfeld (University of Hannover), Jan Louis (University
of Halle), Dieter L\"ust (Humboldt-University Berlin), K. Miesel
(University of Potsdam) and Hermann Nicolai (Albert Einstein Institut (AEI)).
It was the second meeting of this kind in Europe after Amsterdam in 1997
and enjoyed a record participation number of 380 scientists.

From the scientific point of view, the meeting was one of the more quiet 
ones 
with no major breakthroughs reported during the conference, in contrast to 
the previous one at the ITP, Santa Barbara, where the $ADS/CFT$ 
correspondence was 
celebrated. The majority of contributions was rather technical in nature
reporting about small progress in various directions.

However, from the political
point of view the meeting was even more successful than one could have
even dreamt. The local (Berlin, Potsdam and Brandenburg), national
(Frankfurter Allgemeine Zeitung (FAZ), S\"uddeutsche Zeitung etc.) and 
international (Times, New York Times etc.) press reported about the 
conference during the entire week, 
local, national and international broadcasting stations interviewed
several scientists mostly from the AEI and it made it even to
the first national TV company (ARD) of Germany spending ten percent of its 
entire time to report about the conference. Furthermore there was a press 
conference
between reporters from all national TV companies and Micheal Green,
Stephen Hawking, Hermann Nicolai and Edward Witten as well as public
lectures given by Bernard Schutz, Stephen Hawking and Edward Witten
explaining their fields of research to public people, students and 
TV companies. The amount of people that wanted to listen was twice as large
as the number of seats. The public interest in quantum gravity,
especially by young people, was much higher than to be expected by the 
declining physics student enrollment numbers in Germany. 
This week, there is a title page photograph of participants and a 
corresponding report of 
the conference in the ``Spiegel", a weakly political magazine which is 
available throughout the world.

The amount of noise and light that this conference created in the 
national and international press is outstanding. It is hoped that there
will be a positive backreaction for at least string theory-, high 
energy physics- and quantum gravity related physics science throughout 
at least Europe and Germany in particular. That this happened was no 
coincidence : The main organizer of the conference, Hermann Nicolai,
took action long before the conference started by informing press,
broadcasting and TV stations. On the other hand, that they would pick it up
with this amount of enthusiasm was still unexpected.

Coming back to science, the main topics of the conference can be subdivided
roughly into the following headlines :

i) {\it Departure from Supersymmetry}

One of the main drawbacks of string theory is that at present there is only
a perturbative description, apart from the existence of D-branes which are 
conjectured 
to be fundamental excitations at strong string coupling while they are
solitons at low string coupling. These objects were famously used in
stringy black hole calculations in order to probe quantum black holes
(strong coupling) at low coupling. This works well, however, only for 
BPS-D-branes, that is, D-branes which break half of the supersymmetry and 
therefore low coupling calculations in such a setting are protected 
against perturbative corrections through supersymmetry. In practice, this 
means that in string theory one can trust entropy calculations only for
extremal (Reissner-Nordstr{\o}m) black holes. There is an effort therefore
to generalize this rather non-physical restriction.

Sen, Schwarz, Gaberdiel and Horava reported about constructions of stable
non-BPS D-branes, that is, D-branes which do not saturate the Bogomol'ny
bound. This can be done by considering string theories declining from the 
usual world sheet actions with Dirichlet boundary conditions but 
GSO projections different from the usual ones. The result are
Type 0 theories which
are not supersymmertic but contain stable (lightest) D-brane states and
enable one to test string dualities beyond BPS configurations.

Related were considerations by Klebanov, Blumenhagen, Sagnotti and 
Kachru who considered D-brane anti-D-brrane pairs. By employing
the Omega projection and using tachyon condensates one removes the 
instabilities from the theory. 

ii) {\it Brane Worlds}

Famously, the low energy effective action of the closed superstring theories
in absence of D-branes contain the Einstein Hilbert term in leading order of 
the string tension.
Antoniadis, Bachas, Ibanez, Ovrut and Verlinde repeated the analysis that
led to this result in the presence of D-branes and described possible
scenarios that might lead to sub-mm corrections to classical general 
relativity and might solve the problems of coupling constant unification,
stability of the proton and others. Verlinde used ``warped 
compactifications'' in order to explain the smallness of the cosmological 
constant. His analysis suggests that 4d gravity couples to matter on the 
D3 branes and that there is a close relationship between the 5d Einstein 
equations and the 4d renormalization group.

iii) {\it Black Holes}

DeWit described that, if one incorporates higher derivative terms into
the computation of the black hole entropy from black p-branes and uses
Wald's correction, that one can in fact remove the discrepancy between
the quantum statistical entropy and the classical one that was found earlier
in certain scenarios like type IIA on a Calabi-Yau 3-fold.

Hawking considered the stability of rotating black holes in AdS spacetimes 
and found that Kerr black holes are stable.

A very interesting talk was given by Horowitz who considered quasinormal 
modes of evaporating stringy black holes. A mysterious result is that if one 
plots the imaginary part of the lowest frequency quasinormal mode against 
the Schwarzschild radius one finds a linear 
relation with a coefficient  
that equals up to 0.2 percent of accuracy the critical exponents that 
Choptuik found 
for his critical behaviour analysis of classical black holes with the 
same matter content. An interpretation is currently not available.

Peet proved a no-hair theorem for stringy black holes corresponding to
D-p-branes with $p\le 1$.

iv) {\it Renormalization Group Flows}

Gubser and Warner argued that a certain one parameter family 
of diffeomorphisms in the bulk with parameter $r$ 
supergravity theory on an AdS background corresponds to a renormalization 
group flow in the Super-Yang-Mills boundary CFT in the AdS/CFT 
correspondence. Using this one-parameter family of diffeomorphisms they 
show that for AdS-metrics of the form $ds^2=dr^2+e^{A(r)} dX^2$ where 
$dX^2$ is the four-dimensional Minkowski metric one has $A^{\prime\prime}
\le 0$. This suggests that $C=1/(A'(r))^3$ counts the number of degrees of 
freedom available to the boundary QFT at the energy scale 
dual to $r$ and that the exppectation value of the energy momentum tensor 
scales as $C$ ($C$-theorem).

v) {\it Wilson Loops}

Ooguri and Sonnenschein considered the problem to compute the vacuum
expectation value of the Wilson loop operator in the boundary quantum field 
theory 
of the AdS/CFT correspondence. Ooguri found that if one translates the 
BPS condition of D-branes in the bulk into the boundary theory then one
obtains restrictions on the shapes of the loop. For such a special loop one 
can actually compute the Wilson loop operator 
expectation value and finds that it is finite at least for large $g^2 N$ 
where $N$ is the $N$ of the $SU(N)$ gauge group. Sonnenschein found
a finite L\"uscher term in the Wilson loop expectation value which describes 
the quantum fluctuations of the string and one might speculate that these 
fluctuations are therefore finite.

vi)  {\it Non-commutative Gauge Theory}

With great interest people waited for the talks by Seiberg and Witten
who considered open string theory with a constant non-vanishing 
anti-symmetric tensor field. One finds that in the limit of vanishing
string tension and longitudinal components of the target space metric
the effective action is described most easily to all orders in $\alpha'$
by a Yang-Mills theory in the given background spacetime where the 
gauge field components take values in a non-commutative algebra rather 
than in the complex numbers. This description has many technical advantages, 
for instance in the description of instanton moduli spaces as described 
by Nekrasov and Schwarz.

vii) {\it Others}

Staudacher computed Matrix model expectation values by Monte Carlo methods
indicating that certain integrals seem to be finite although the matrix
model potantial has flat directions. Dijkgraaf and Maldacena reviewed 
topics in 
the AdS/CFT correspondence. Polchinsky reported about progress in the 
DLCQ (discrete light cone quantization) of matrix theory and SYM.
Hoppe and Yoneya described open problems in the matrix model approach to 
M-Theory. Douglas, Mayr, Townsend and Karch reviewed various aspects of
supersymmetric p-brane actions. Bousso and Gibbons talked about holography
indicating that the correpondence between bulk and boundary theories 
works only in special cases as for example if the background spacetime is 
AdS. Kallosh and Berkowits tried quantization of superstrings in 
AdS spaces (using BRST and twistors) and Ramond-Ramond backgrounds 
respectively. Green and Obers chose instanton matching as their topics
and Theisen gave new insights into gravitational and conformal anomalies.

The conference ended in a summary by David Gross who stated that he 
missed talks that :\\
$\bullet$ tested the AdS/CFT correspondence beyond $N=\infty$,\\
$\bullet$ gave a spacetime description of black hole formation and 
evaporation,\\
$\bullet$ gave a dual or holographic description of flat space and 
cosmology and\\
$\bullet$ gave a non-perturbative and background independent formulation 
of ?-theory.\\
He ended the conference with a ``hello again" in Michigan next year and 
probably the Tata institute, Mumbai in 2001.

\end{document}